\documentclass[english,pre,superscriptaddress,twocolumn]{revtex4-2}
\usepackage{textcomp}
\usepackage[utf8]{inputenc}
\usepackage{babel}
\usepackage{units}
\usepackage{amsmath}
\usepackage{stackrel}
\usepackage{graphicx}
\usepackage{orcidlink}
\usepackage{esint}
\usepackage{xcolor} 
\usepackage{amssymb} 
\usepackage{hyperref}
 \usepackage[normalem]{ulem}

\makeatletter

\newcommand{\lyxmathsym}[1]{\ifmmode\begingroup\def\b@ld{bold}
  \text{\ifx\math@version\b@ld\bfseries\fi#1}\endgroup\else#1\fi}

\providecommand{\tabularnewline}{\\}
\newcommand{\lyxdot}{.}

\usepackage{units}

\usepackage[version=4]{mhchem}

\makeatother

\begin{document}
\title{Material and thermal properties of MgCl$_{2}$ molten salt by ab initio
and machine-learning molecular-dynamics simulations}

\author{Roberto Llovera\orcidlink{0009-0006-1853-0925}}
\email{robertollovera@cnea.gob.ar}
\affiliation{Departamento de Física de Materia Condensada, GIYA, CAC-CNEA, Av. Gral. Paz
1499, San Martín, Pcia. de Buenos Aires, Argentina}
\affiliation{CEMA, Laboratorio de Metrología de Radioisótopos, CAE-CNEA, Av. Presbítero
Juan Gonzalez y Aragón 15, Ezeiza, Pcia. de Buenos Aires, Argentina }

\author{María Andrea Barral,\orcidlink{0000-0001-7670-2664}}
\affiliation{Departamento de Física de Materia Condensada, GIYA, CAC-CNEA, Av. Gral. Paz
1499, San Martín, Pcia. de Buenos Aires, Argentina}
\affiliation{Instituto de Nanociencia y Nanotecnología, INN-CONICET-CNEA}

\author{Verónica Vildosola,\orcidlink{0000-0002-7412-516X}}
\affiliation{Departamento de Física de Materia Condensada, GIYA, CAC-CNEA, Av. Gral. Paz
1499, San Martín, Pcia. de Buenos Aires, Argentina}
\affiliation{Instituto de Nanociencia y Nanotecnología, INN-CONICET-CNEA}

\author{Florencia  Cantargi}
\affiliation{Asuntos Académicos, Gerencia de Área Articulación Institucional, CNEA}

\author{Claudio Pastorino,\orcidlink{0000-0002-4833-5999}}
\email{claudiopastorino@cnea.gob.ar}
\affiliation{Departamento de Física de Materia Condensada, GIYA, CAC-CNEA, Av. Gral. Paz
1499, San Martín, Pcia. de Buenos Aires, Argentina}
\affiliation{Instituto de Nanociencia y Nanotecnología, INN-CONICET-CNEA}


\begin{abstract}

We study the structural, thermophysical and dynamical properties of 
molten MgCl\textsubscript{2} over a wide temperature range using Ab Initio Molecular Dynamics (AIMD) and molecular dynamics simulations based on machine-learning interaction potentials (MLIPs). 
MLIPs can achieve near-AIMD accuracy at a fraction of the computational  cost, enabling simulations of larger systems and longer timescales.
This opens up the possibility of studying the system under out-of-equilibrium conditions, which in turn allows for the calculation of physical properties, such as the thermal conductivity or viscosity,  that cannot be obtained reliably for the typical time and length scales accessible to AIMD.
We follow two complementary approaches to develop and evaluate the MLIPs. Firstly, we develop a Deep Potential (DP) from scratch using AIMD trajectories at different temperatures.
Secondly, we also evaluate two  out-of-the-box foundation models and a fine-tuned version of one of them. The fine-tuning is performed using  configurations from the AIMD trajectories originally used to train the Deep Potential.
We compare the predictions of the trained DP, the foundation models and AIMD simulations  with the available experimental data and provide a comprehensive calculation of the most important physical quantities of molten MgCl\textsubscript{2}.
We compute densities, radial and angular distribution functions,  thermal conductivity, heat capacity, viscosity and diffusion coefficients over the 1000-1600~K temperature range. This range is relevant to important technological applications such as thermal energy storage,  generation IV nuclear reactors and concentrated solar power systems. Finally, we briefly review the accuracy and performance of the different MLIPs versions used throughout the work. 




\end{abstract}
\maketitle

\section{Introduction}



Molten MgCl\textsubscript{2} salt has applications as a thermal energy storage (TES) or as a heat transfer fluid (HTF) \citep{Caraballo2021} in developments like Gen 3 concentrated solar power (CSP) generators \citep{CSP} and Gen IV molten salt reactors (MSR) \citep{MSR}. For these purposes it has a typical operational temperature range that is defined by the mixture it belongs to and by the composition of the given mixture of molten salts. In its pure form, molten MgCl\textsubscript{2} salt has a temperature range between 987~K and 1685~K. Composite chloride-based molten salts have features that make them suited for high-temperature applications.
Among those features is the low melting point of these mixtures, which in its  eutectic compositions,  produce the lowest possible melting point for the mixture. The high working temperature range also poses other benefits, like higher thermal conductivities and lower viscosities. Optimizing these and other thermophysical properties   constitutes an important goal for the design and daily operation of CSP and MSR facilities.
Thermal stability of these materials at +700~C temperature is another desirable feature, because it avoids the degradation of the salt while being used continuously. Economic convenience is also important when planning the use of molten salts. Chloride-based salts are cost-effective as compared to nitrate-based salts, making them appealing for research studies in this field. A disadvantage of these salts is their significant corrosion on metallic alloys. This is a serious problem for the metallic pipes transporting  the salts in real-life facilities. Modifying these salts and  the alloys used  in the pipes in order to make these effects less damaging is currently an active area of research \citep{PATANGE2026114090,XIAO2025113531}. Performing experimental measurements of molten salt properties is a difficult task at high temperatures due to corrosion effects and, in some cases (as in FLiBe), because of the toxicity of the substances involved. This is one of the reasons computational modeling of molten salts has evolved so much through the years. Sharma et al. \citep{Sharma2021} comment on  difficulties and particularities of X-ray and neutron diffraction of  molten salts, in comparison with ionic liquids.

In this context, it is difficult to obtain precise information on thermophysical and structural properties of pure molten salts and their mixtures. 
Online databases exist, like ORNL Molten Salt Database \citep{ORNL_DB}, NIST Phase Equilibria Diagrams Database \citep{NIST-database} or TCSalt \citep{TCSalt} that gather a significant amount of information from  experiments and simulations, but many properties are missing or  incomplete in the operational temperature range of the applications of interest.
Several experimental studies measure thermodynamic and transport properties of molten MgCl\textsubscript{2} that provide essential information for general reference. 
Janz \citep{Janz1988} measures density, surface tension, electric conductance and viscosity of MgCl\textsubscript{2} as a part of a long-standing experimental effort that is a reference for molten salts properties. Parker et al. \citep{Parker2022} measure the melting temperature, the enthalpy of fusion, density (up to 1300~K) and vapor pressure (up to 1500~K). Density, expansion coefficient and electrical conductivity is measured by Sato et al. \citep{Sato1999} as a part of their study of the MgCl\textsubscript{2}-CaCl\textsubscript{2} binary mixture. Heat capacity is measured by Moore \citep{Moore1943} in 1943, thus providing one of the earliest values of this thermodynamic property for MgCl\textsubscript{2}. Structure factors and pair correlation functions are investigated experimentally by Biggin et al. \citep{Biggin1984} using neutron diffraction on isotopically enriched samples. Thermal conductivity is studied experimentally in the works of Bystrai \citep{bystrai1975thermal} and Filatov \citep{Filatov}, showing a remarkable difference with other scientific sources. Golyshev et al. \citep{Golyshev1992} and Perry et al. \citep{Green2019Perry} present thermal conductivity values near the melting point of MgCl\textsubscript{2}. Viscosity is also studied experimentally by Tørklep and Øye \citep{Toerklep_1982} using a viscosimeter that measures the damping of the oscillations of a noble-metal cylinder. In this method, no need for calibration against reference liquids is needed. It is important to mention that many of the experimental studies  do not span the complete temperature range of molten MgCl\textsubscript{2}.  This is one of the motivations behind the choice of the 1000-1600~K temperature range in this work.

Complementary computational studies investigate the  properties of molten MgCl\textsubscript{2} using a variety of simulation methods. Attarian et al. study NaCl-MgCl mixtures with various  MLIPs generated with ACE and an active learning procedure \citep{ATTARIAN2025113409}.
They also study the limiting case of MgCl\textsubscript{2} and calculated density,  radial distribution functions and viscosity.
Gheribi et al. propose a thermodynamic model for the thermal conductivity of various molten salts including MgCl\textsubscript{2} \citep{Gheribi2014} and, in a follow-up work, a further refinement of the model, especially for MgCl\textsubscript{2}, using PIM potentials  and AIMD simulations \citep{Gheribi2022}, with the goal of achieving prediction capability of molten salt mixtures for Concentrated Solar Thermal Power systems.
Density, diffusion coefficient and thermal conductivity  of MgCl\textsubscript{2} are obtained with a deep potential, generated with DP-Gen by Liang et al. \citep{Liang2020}. They also compare their results with experiments and AIMD data, in a somewhat narrower temperature range than that studied here. Lu et al. study the thermal properties of the KCl-MgCl\textsubscript{2} eutectic mixture with MD simulations based on the Born-Mayer-Huggins potential \citep{Lu2021}. They calculate radial distribution functions, coordination properties, shear viscosity,  and thermal conductivity as a function of temperature and  relative composition of both salts. Duemmler et al. study thermophysical properties of molten NaCl\textsubscript{2}-MgCl\textsubscript{2} mixture by AIMD for the eutectic and as a function of composition \citep{Duemmler2022}. In a second paper \citep{Duemmler2023}, the authors explore dynamics properties of chloride molten salt and in particular, compute by AIMD, the diffusion coefficients at 1000~K, and viscosity in the 1000-1300~K temperature range. 


Computer simulations have become a promising approach for exploring physicochemical properties as functions of salt concentration and temperature {\sl in silico}, before conducting experiments. However, obtaining accurate and consistent results remains a significant challenge \citep{Kalita2025}.

Quantum level accuracy is possible for the prediction of thermal and structural properties by performing Ab Initio Molecular Dynamics simulations.
The downside of this technique is that the trajectories can be obtained only on  systems of a few hundred atoms, evolved for a few dozen picoseconds, due to the high computational cost and poor scaling in system size, inherent to Density Functional Theory (DFT) calculations.
The development of MLIPs has given a big impulse to the study of properties of complex systems such as these salts. A significant reduction in computational cost allows for the calculation of many properties with quantum-level DFT precision  in systems with thousands of atoms, evolving up to nanoseconds. 

MLIPs can be developed from scratch with a careful training of the  neural network by using  DFT configurations,  targeted to the systems under study \citep{Zhang2018,Behler_2007,Behler2017,Zhang2022,Wang2018,Lu2022,Zeng2025} or,  alternatively, 
using the  so-called MLIPs  foundation models. The latter  can span over most of the Periodic Table
and a relatively wide range of  physicochemical environments, providing a good starting point for the study of complex systems. 
These models have been trained on databases of  hundreds of thousands of configurations from Ab Initio calculations of a wide variety of systems. They allow  the effective  harvesting of a huge amount of computational time to pack the interaction information among atoms in a single foundation MLIP \citep{Batatia2025,Kovacs_MACE-OFF,SuperSalt_Shen2025,CHGnet_Deng2023}.
Foundation models can be applied out-of-the-box and if the precision obtained for the physical properties have  a high error as compared to AIMD, they can be fine-tuned with new DFT data to improve accuracy \citep{Batatia2025}.

In this work we present an extensive set of  thermodynamic, structural and transport properties of molten MgCl\textsubscript{2} salt.
We use AIMD simulations  to  develop and test a Deep Potential \citep{Zhang2018} and employ MACE-foundation-model simulations \citep{Batatia2025} to compare with our DP model and the available experimental data. 

We study the density, the radial distribution function, the coordination number, the angular distribution function, heat capacity, thermal conductivity, viscosity and ionic diffusion for temperatures in the 1000-1600~K range. A comparison of these methods  is presented in order to test the accuracy of DPMD and MACE simulations with AIMD simulations.  This systematic assessment provides an insight into the capabilities and limitations of system-specific and foundation machine-learning potentials for predicting the properties of molten MgCl\textsubscript{2}.
Our simulations aim at covering the extreme case of pure MgCl\textsubscript{2} so that a comparison of molten mixtures against pure chlorides, for most of the structural, thermodynamic and transport properties is possible.
It also contributes to  models of molten salt mixtures \citep{Gheribi2022}, that  depend on pure molten salt properties to predict the physical behavior of binary or ternary mixtures.

\section{Computational methods}

\subsection{\label{subsec:Ab-initio-Molecular-Dynamics-1}Ab Initio Molecular
Dynamics}

 We first carry out a set of AIMD simulations in the canonical (NVT) ensemble. The density functional theory calculations are performed within the projector augmented-wave  (PAW) formalism, as implemented in the Vienna Ab initio Simulation Package (VASP) \citep{Kresse_1993,Kresse_1996,Kresse_1996b}. 
We use the general gradient approximation by  Perdew-Burke-Ernzerhof (PBE) \citep{PhysRevLett.77.3865}  for the exchange and correlation functional of the electronic interactions. The valence configurations considered for the pseudopotentials are (s2p0 13Apr2007) for Mg  and (s2p5 06Sep2000) for Cl. The energy cutoff for the plane wave expansion of the electron wave functions is set at 350~eV, thus providing a balance between computational cost and accuracy.
Given the large size of the simulation cell, only the $\Gamma$ point is used for Brillouin-zone sampling. We consider Grimme dispersion interactions through the DFT-D3 correction \citep{Grimme_2010,Grimme_2011}. 

 As a first step, we perform a series of NVT simulations at different volumes for each temperature. The calculated pressure–volume data are fitted to the Birch–Murnaghan equation of state \citep{Birch_1947,Murnaghan_1944} to determine the equilibrium density.
 We use a Nosé-Hoover thermostat and a mass parameter set to 1 (corresponding to a thermostat oscillation period of approximately 130 fs).
  Pressures are calculated with 20~ps runs, which is an
appropriate total time for the evolution of the system, because thermodynamic equilibrium is reached at about 2~ps.
 A sample of 36 Mg and 72 Cl atoms is initially arranged in a cubic cell.
 The time step
of the simulation is set to 1 fs, in order to correctly track the
motions of the atoms. The calculated equilibrium
density values define the conditions under which the rest of the structural
and thermodynamic properties are calculated.

Using these equilibrium density values at different temperatures, we
conduct 40~ps AIMD simulations, in order to calculate the radial distribution function, the coordination number, the coordination number distribution, the angular distribution function and heat capacity of molten MgCl\textsubscript{2}.
The length of these simulations is chosen to obtain a long enough trajectory over the phase space of the system while simultaneously generating a representative set of atomic configurations for the training of the machine-learning interaction potential.
The AIMD results for the structural properties are presented in Sections \ref{subsec:Radial-distribution-function}, \ref{subsec:Angular-distribution-function} and \ref{subsec:Heat-capacity-1}.
In the following paragraphs we provide some details on the calculation of the studied properties.

\subsubsection{Radial Distribution Function (RDF)\label{subsec:Radial-Distribution-Function}}

This function provides a measure of the probability of finding an
atom of type $\beta$ within a spherical shell of size
$dr$ at a distance $r$ from an atom of type \(\alpha\), relative
to an ideal gas at the same density and temperature. It is defined
by the equation:

\begin{equation}
g_{\alpha\beta}(r)=\frac{\left\langle N_{\alpha\beta}(r)\right\rangle }{\frac{4\pi}{3}((r+dr)^{3}-r^{3})N_{\alpha}\rho_{\beta}}\label{eq:1}
\end{equation}

\noindent where $\rho_\beta$ is
the density of type $\beta$ atoms, $N_\alpha$
is the number of type $\alpha$ atoms and $<N_{\alpha\beta}(r)>$
is the ensemble average of the number of type $\beta$
atoms around a type $\alpha$ atom at distance $r$. This
ensemble average is equivalent to a temporal average over the
trajectory of the simulation. Once the RDF is obtained, it can be
integrated to obtain the coordination numbers $N_{\alpha\beta}$
from the equation:

\begin{equation}
N_{\alpha\beta}=\sideset{}{}\intop4\pi r^{2}\,\rho_{\beta}\,g{}_{\alpha\beta}(r)\,dr\label{eq:2}
\end{equation}

The coordination number of the $\beta$ atoms around
the $\alpha$ atoms can be obtained by integrating from
$r=0$ to $r=r_{min}$, where $r_{min}$ is
the radial location of the first minimum of the RDF, as shown in Section
\ref{subsec:Radial-distribution-function}.

\subsubsection{Angular Distribution Function (ADF)\label{subsec:Angular-Distribution-Function}}

The internal structure of MgCl\textsubscript{n} complexes within molten MgCl\textsubscript{2} salt can be studied using the angular distribution function (ADF) of atomic triplets where the center is a Mg\textsuperscript{2+} ion. 
In general, the ADF is defined by the following equation

\begin{equation}
\theta_{\alpha\beta\gamma}=\left\langle \cos^{-1}\left(\frac{r_{\alpha\beta}^{2}+r_{\alpha\gamma}^{2}-r_{\beta\gamma}^{2}}{2\,r_{\alpha\beta}\,r_{\alpha\gamma}}\right)\right\rangle \,,\label{eq:3}
\end{equation}

\noindent where \( r_{ij} = \left\lvert {\bf r}_i - {\bf r}_j\right\rvert\) for \(i,j=\alpha,\beta,\gamma\). The ADF gives information on the angle formed by the $\alpha-\beta-\gamma$ triplet of ions. It is a measure of the average deformation of the molten salt complexes (e.g., [MgCl\textsubscript{4}]\textsuperscript{2-}, [MgCl\textsubscript{5}]\textsuperscript{3-}) for a given temperature. We choose the Cl-Mg-Cl triplet, where the
Mg\textsuperscript{2+} cation is at the center of the triplet and measure the distances between Mg\textsuperscript{2+} and Cl\textsuperscript{-}
ions up to the first minimum of the Mg-Cl RDF.

\subsubsection{Heat capacity\label{subsec:Heat-capacity-2}}

Heat capacity 
is
proportional to  the variance of the total energy of the system,
which is expressed in the equation

\begin{equation}
C_{V}\,[\nicefrac{J}{(mol\,K)}]=\frac{1}{n\,k_{B}\,T^{2}}\,{\rm var}(E), \label{eq:heat_capacity}
\end{equation}

\noindent where, $n$ is the number of moles in the simulation cell, $k_{B}$ is the  Boltzmann constant and
${\rm var}(E)$ is the energy variance. This property is difficult to obtain accurately by AIMD, due to its usually high fluctuations which should be averaged out with a small MD sample and a relatively short time evolution. In this sense, this is a property which would
greatly benefit from ML-generated interaction potentials, since these
MD simulations allow for property calculation on larger systems  and longer trajectories. The results obtained by DPMD are presented in Section \ref{subsec:Heat-capacity-1}.

\subsection{\label{subsec:Deep-Potential-generation-1} Generation of the Deep Potential}

Some of the configurations obtained from the AIMD trajectory in the
NVT ensemble are used to train the Potential Energy Surface (PES)
which determines the interaction forces between ions in the DPMD simulation. We use the DeepMD-kit package \citep{Zeng2025} to generate a machine-learned interaction model from scratch, by using 
configurations from AIMD trajectories at different temperatures.
We perform AIMD simulations at seven different temperatures from
1000~K to 1600~K at intervals of 100~K and we select 70000 configurations
randomly. DeepMD-kit performs the PES generation with the input configurations
through a three-step process: a \emph{training stage}, a  \emph{validation
stage} and, finally, a \emph{model testing stage} of the Deep Neural
Network (DNN). A \emph{loss function $L$} is minimized after a number
of training steps. The loss function is defined in the following manner:

\begin{equation}
L(x;\theta)=p_{E}\,L_{E}(x;\theta)+p_{F}\,L_{F}(x;\theta)
\end{equation}
 
\begin{equation}
L_{E}(x;\theta)=\frac{1}{N}(E(x;\theta)-E^{*})^2
\end{equation}
\begin{equation}
L_{F}(x;\theta)=\frac{1}{3N}\stackrel[k=1]{N}{\sum}\stackrel[\alpha=1]{3}{\sum}(F_{k,\alpha}(x;\theta)-F_{k,\alpha}^{*})^{2}\,.
\end{equation}

Here, $L_E$ measures how far  the DFT energies
of the training configurations are from the values obtained by the
DP model, after some training steps. $L_F$ represents  the same measure
 for the atomic forces. In the equations above, $x =\{x^{k}\}$
indicates the dataset used for training, while $x^{k}
= \{x_{1}^{k},\dots,x_{N}^{k}\}$
is a single data frame composed of all the degrees of freedom of the
system. $\theta$ represents all the weights and biases
of the DNN, which will be modified in the minimization process. $p_E$
and $p_F$ are prefactors for the energy and the forces respectively, that can
be set to initial and final values by the user. These prefactors, which are functions
of the training step, weight differently the loss functions of energy
and force in the different stages of the training process. 56000 AIMD configurations are randomly chosen for the
training stage, 7000 for validation and 2000 for model testing. The
loss function for the validation data is calculated as training proceeds. Validation configurations are used to have a measure of the ``overfitting'' that may occur while training \citep{Behler2017}. The testing configurations
are used to calculate the final root mean square errors (RMSE) metrics on a completely new dataset, not used at all during the training stage.
The training stage consists of 200000 training
steps. The prefactors were set to the following initial and final
values: $p_{E,\rm{initial}} = 0.02$, $p_{E,\rm{final}}=1$, $p_{F,\rm{initial}} = 1000$ and $p_{F,\rm{final}}=1$. This choice is done so that the minimization process produces a PES that mainly fits the atomic forces better than the energies.
Unlike early ML models, DeepMD-kit adaptively builds the variables which
describe the atoms of the system and their closest environment during
the PES generation. It uses flexible \emph{atomic environment descriptors}
instead of the atom positions to achieve this goal. These descriptors
are obtained through the use of an \emph{embedding} DNN, which is
in contrast to other type of descriptors, usually fixed before the
start of the training stage. This is the case, for example, for  the Behler-Parrinello symmetry functions \citep{Behler_2007}. In DeepMD-kit, the definition of the atomic environment descriptors is an integral part of the training process. We use the model descriptor \emph{se\_e2\_a} which takes into consideration both the radial and angular information between a given ion and all the other ions within its close spherical environment
up to a cutoff radius. We set this cutoff to 8~Å for our PES. The
embedding network is composed of  three layers with \(25\), \(50\) and \(100\) nodes per layer, respectively. The \emph{energy fitting network}, which relates the atomic environment descriptors with the total energy of each system
configuration, has three layers with \(240\) nodes in each of them. The
weights and the biases of both DNNs are initialized with random numbers. This approach simultaneously optimizes the variables that describe the atomic environments and the energies and forces produced by the ML model, in order to approximate them as closely as possible to the DFT data. After generating the ML model as described above, it is compressed with the \textit{compress} command of DeepMD-kit, in order to enhance the performance of the PES evaluation for forces and energies within the molecular-dynamics simulation \citep{Lu2022}.

\begin{figure}
\centering{}\includegraphics[width=0.98\columnwidth]{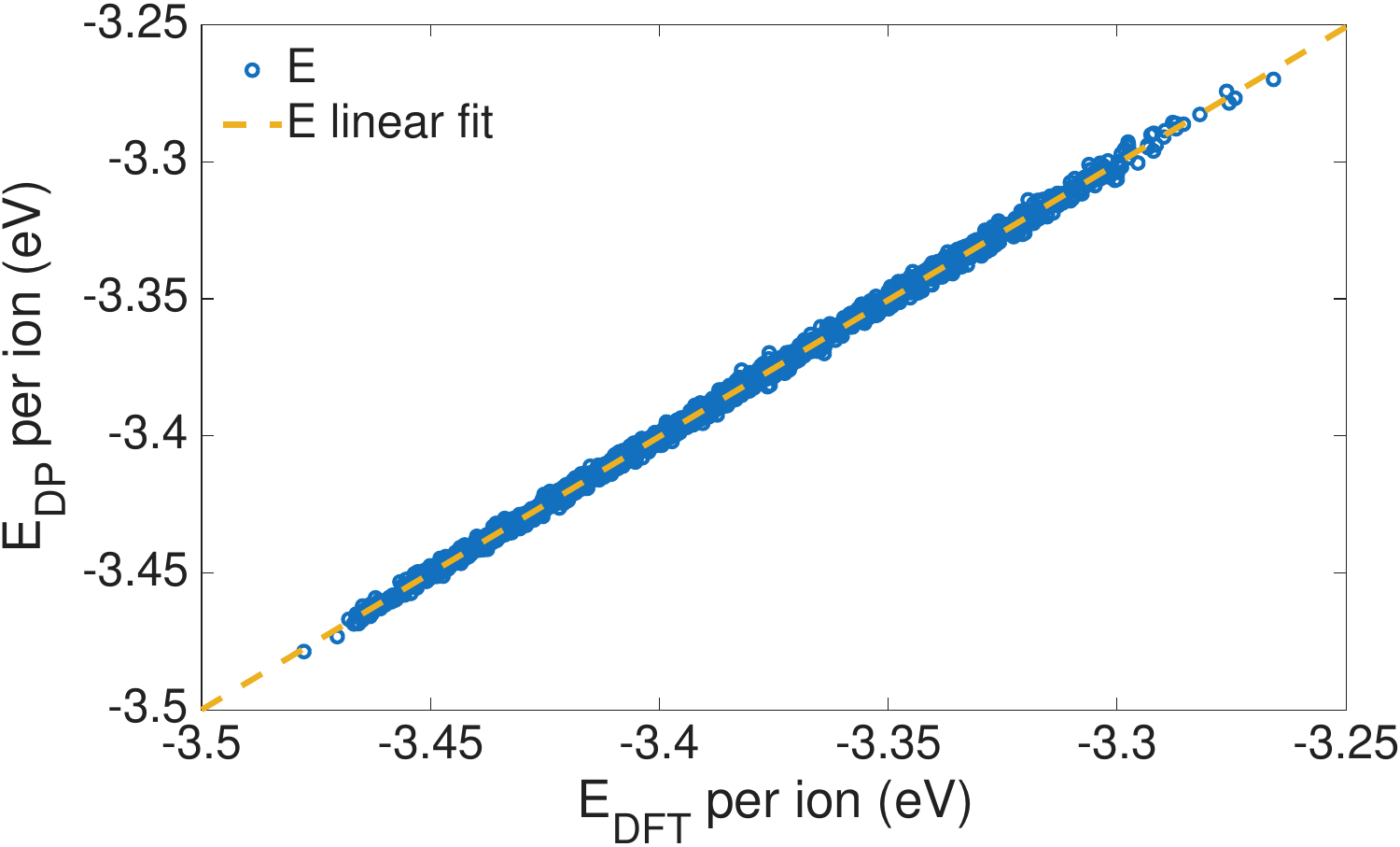}\vfill{}
\includegraphics[width=0.98\columnwidth]{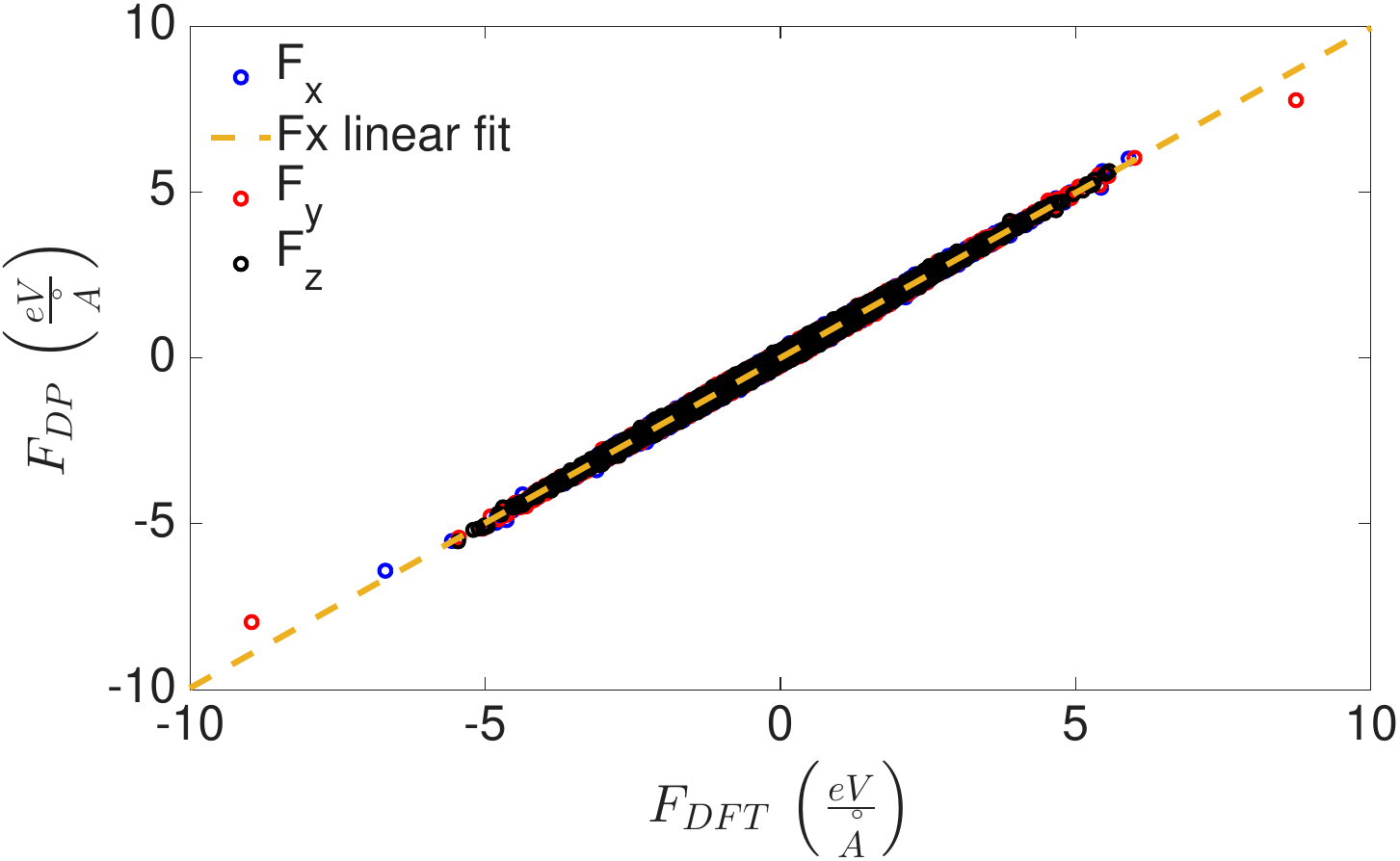}\caption{\label{fig:Parity-plot-E}
Parity plots of energy per ion (upper panel) and forces (lower panel) for the generated molten MgCl\protect\textsubscript{2} deep-potential model, in the temperature interval of 1000--1600~K.}
\end{figure}

Figure \ref{fig:Parity-plot-E} shows a so-called parity plot in which
the energy per ion obtained by the evaluation of the Deep Potential
on the testing dataset is plotted against the energy obtained by the DFT calculation of AIMD.
The linear fit of \(E_{\rm DP}\) vs. \(E_{\rm DFT}\) is given by

\begin{equation}
E_{DP}=0.9967\,E_{DFT}-0.0115
\end{equation}

\noindent with an energy RMSE of \(2.39\,meV\). The
slope close to 1 indicates that the training produces a DP model for the PES that accurately reproduces the DFT energies of the complete system.

This is very important because the interaction forces between atoms
are calculated directly from the differentiation of the configurational
energy \(E\) with respect to the instantaneous interatomic positions:

\begin{equation}
F_{k,\alpha}=-\frac{\partial E}{\partial r_{k,\alpha}},
\end{equation}

\noindent where \(k\) labels the particles and \(\alpha\) represents the cartesian coordinate. The lower panel of Figure \ref{fig:Parity-plot-E} shows the parity plots which compare the cartesian components of the forces on atoms  for  DFT and DP using the testing dataset. The linear fit of \(F_{x, {\rm DP}} \) vs. \(F_{x, {\rm DFT}}\), for example, is given by:

\begin{equation}
F_{x, {\rm DP}}=0.9948\,F_{x, {\rm DFT}}-1.0\times10^{-6}
\end{equation}

\noindent with a force RMSE of \(5.8\times10^{-2}~eV/\)\AA. Similar
fits are obtained for $F_{y}$ and $F_{z}$.

The slope of the fit of this curve is very close to $1$ and it also
has a negligible offset. This evidences the high accuracy that is obtained for
the calculation of the atomic forces from the DP-generated PES. We present
more results about the training procedure of the ML model (including model deviations for an ensemble of four models) in Section II of the Supplementary Material.

\subsection{\label{subsec:Deep-Potential-Molecular-1}Deep-Potential Molecular
Dynamics}

The MLIP model, generated with the process described in the
previous section, is used as the interatomic potential for the molecular dynamics simulations of molten MgCl\textsubscript{2} at different temperatures.
As shown in Section \ref{subsec:Deep-Potential-generation-1}, this
model retains an accuracy that is very close to that
of the DFT data, with about \(4000\times\) of  speedup in computation time (see Section \ref{sec:performance_MLIPs} for details on performance).  This allows for the simulation of larger systems and longer trajectories, which
is needed for the appropriate calculation of many physical quantities.
In the case of the thermal conductivity and heat capacity, we run
simulations with up to 4320 atoms for 12\, ns and 10\, ns, respectively.
This represents  a significant contrast to AIMD, where only a few hundred atoms and evolutions up to tens of picoseconds can be achieved in reasonable computing times. We use the  LAMMPS package to perform the molecular-dynamics simulations with this MLIP potential \citep{Thompson2022}.

\subsubsection{\label{subsec:Thermal-conductivity-1}Thermal conductivity}

For the calculation of thermal conductivity, we
use the Reverse Non-Equilibrium Molecular Dynamics (RNEMD) method by Müller-Plathe \citep{MuellerPlathe1997}.
At each given temperature, we calculate the size-independent
thermal conductivity, that is, the thermal conductivity corresponding to an infinitely large system, as explained in the following paragraphs. We create four systems of different sizes for each temperature with their corresponding equilibrium density.
 The number of ions in each system are set to  \(1728\), \(2592\),
\(3456\) and \(4320\), respectively. The simulation cells for each size are set to the volumes \( 2L\times 2L \times D' \), with
\(D'=\{4L,6L,8L,10L\}\).  \(L\) stands for the cell length used for the AIMD simulations. The MD cell is divided into \(20\) slabs (or bins) across \textit{z} direction. The slabs, numbered as \(1\) and \(20\) at the extremes of the MD simulation cell,  are set as the \emph{cold slabs}, while the slab number \(11\), at the center of the simulation cell, is defined as the \emph{hot slab}. Periodic boundary conditions are applied in the three cartesian directions. The system is removed from its
thermodynamic equilibrium state into a stationary non-equilibrium
state by the swap of ion velocities between the cold and hot slabs \citep{MuellerPlathe1997}.
This induces a temperature gradient that can be related with the heat
flux along the \textit{z} direction through the Fourier law:

\begin{equation}
\mathbf{J}=-\lambda\,\mathbf{\nabla}T,
\label{eq:fourier_heat}
\end{equation}

\noindent where \(\lambda\) indicates the thermal conductivity, \textbf{J}
the mean heat flux and $\nabla T$ a constant thermal gradient induced
by the non-equilibrium RNEMD scheme. The heat flux can be calculated
directly from the rate of swapped-particle velocities. For a given
swap time value, the particle velocity of highest kinetic energy in the cold slab is exchanged with the particle velocity of the minimum kinetic
energy of the hot slab. This fixes the energy transfer between slabs
to a known value. The velocity exchanges are performed every \(N_{{\rm swap}}\) time steps.
Energy is conserved within this method, so the NVE ensemble is used for the simulations, with a time step of 1~fs.
A suitable value for $N_{{\rm swap}}$ must be selected, so that the temperature gradient is relatively small to make sure that the system is in linear-response regime, but the number of swaps should be high enough, such that a good overall statistics is obtained for a reasonable trajectory length. 
The dependence of the thermal conductivity on the swap rate will be presented in Section \ref{subsec:Thermal-conductivity}. The  thermal conductivity values are obtained from equation:

\begin{equation}
\lambda=\underset{\nicefrac{\partial T}{\partial z}\rightarrow0}{\lim}\,\underset{t\rightarrow\infty}{\lim}\,-\frac{\left\langle J_{z}(t)\right\rangle }{\left\langle \nicefrac{\partial T}{\partial z}\right\rangle }
\end{equation}

The instantaneous temperature of slab \textit{k} is given by

\begin{equation}
T_{k}=\frac{1}{3n_{k}k_{B}}\stackrel[i\in k]{n_{k}}{\sum}m_{i}\,v_{i}^{2}~,
\end{equation} 

\noindent where \(k_{B}\) is the Boltzmann's constant and \(n_{k}\) indicates the number of atoms in slab \(k\). The final expression for the thermal conductivity in terms of the energy transfers in each velocity swap is given by:

\begin{equation}
\lambda=-\frac{\underset{{\rm swaps}}{\sum}\frac{m}{2}(v_{h}^{2}-v_{c}^{2})}{2tL_{x}L_{y}\left\langle \nicefrac{\partial T}{\partial z}\right\rangle }\,,
\end{equation}

\noindent where \(t\) is the simulation time and $L_{x}$ and $L_{y}$ are both
equal to $2L$. The sub-indices $h$ and $c$ refer to the hot and
cold particles respectively, whose velocities are exchanged.
 
\subsubsection{\label{subsec:Viscosity}Viscosity}

We calculate the viscosity as a function of temperature by the Green-Kubo method. This implies the calculation of the time integral of the average  stress autocorrelation function (SACF), according to the following equation:

\begin{equation}
\eta=\frac{V}{3k_{B}T}\,\sideset{}{_{0}^{\infty}}\intop\underset{\alpha<\beta}{\sum}\left\langle P_{\alpha\beta}(0)\,P_{\alpha\beta}(t)\right\rangle \,dt.
\label{GK-viscosity}
\end{equation}

Here, \(V\) is the total volume of the simulation cell, \(k_{B}\) the Boltzmann constant, \(T\) the temperature and \(P_{\alpha\beta}\) the non-diagonal components of the stress tensor. We set up a cubic simulation cell
with \(4320\) atoms at the proper equilibrium density for each temperature.
The system is brought to thermal equilibrium in an NVT ensemble with a Langevin thermostat with characteristic time of 500~fs. Afterwards, we run several independent simulations of
0.7~ns to calculate the average integral of the SACF. The simulation
interval is chosen so that we get a converged value of viscosity
upon integration. The autocorrelation function is calculated for a maximum time of 7~ps (though we show up to 4~ps for readability's sake in Figure S5 of the Supplementary Material), because the contribution of the SACF time integral to the viscosity $\eta$ is found to be negligible beyond that time.
The mean value and the uncertainty of the viscosity for each temperature is calculated from the converged values of each simulation.

\subsubsection{\label{subsec:Diffusion-coefficient}Diffusion coefficient}

We calculate the diffusion coefficient \textit{D} for the  Mg\textsuperscript{2+} and Cl\textsuperscript{-}
ions. The diffusion coefficient is defined as a function of the mean square displacement (MSD) of each ion in the following equation:

\[
D=\underset{t\rightarrow\infty}{\lim}\frac{\left\langle |\mathbf{\mathbf{r_{\mathrm{i}}\mathrm{(t)-\mathbf{r\mathrm{_{i}(0)|^{2}}}}}}\right\rangle }{6t}
\]

The  \( \langle \cdots \rangle \) symbol represents ensemble averages over a given species. In practice, it is calculated from the fitting of the slope of the MSD vs. time plot for each species, for long enough trajectories to make sure that the system  is in diffusive regime. 
We average over 10 sets of statistically independent configurations  \(\mathbf{r}_i(t=0)\) to obtain an appropriate error estimation  of \textit{D} and subtracted the  motion of the center of mass of the system.

\subsection{\label{subsec: MACE}MACE foundation model}

Recently, the MACE architecture \citep{Batatia2022_mace_seminal} has evolved to such a degree in which it is now possible to use a MLIP which is built on configurations of systems throughout a major part of the Periodic Table and is targeted for use on a wide variety of chemical environments. These models are called ``foundation models''. The main advantage of these so-called  uMLIPs (universal MLIPs) is the possibility of using equivariant ML interaction models with a reasonable out-of-the-box accuracy. Additionally, these models can be fine-tuned with a few DFT calculations of the specific system to improve their accuracy \citep{Batatia2025,Kovacs_MACE-OFF,SuperSalt_Shen2025,CHGnet_Deng2023}. 

We calculate relevant physical properties of  molten MgCl\textsubscript{2} with some uMLIPs models. We use the MACE-MP-0b3 foundation model \citep{Batatia2025}, which is trained with the MPtrj Materials Project dataset \citep{Jain2013} and the MACE-OMAT-MATPES foundation model \citep{Kaplan2025_arxiv}.
In the latter case, we also generate a modified version of the model, using the {\sl multihead replay finetuning} method.
This fine-tuning procedure is designed to prevent the "catastrophic forgetting" that can occur when modifying foundation models without taking into consideration the original dataset from which the foundation model is made \citep{Batatia2025}. 
The basic idea is to capture the features of the new dataset while retaining the properties of the original dataset, by using two ``training heads'', one assigned to the original dataset and another one for the ``fine-tuning'' dataset \citep{Batatia2025}. 
The MACE-MP-0b3 model is built around configurations at 0~K, while the MACE-OMAT-MATPES model also has configurations at 300~K generated by performing MD simulations \citep{batatia2025_mace_matpes_omat,BarrosLuque2026}.
We aim at  studying  how well these foundation models extrapolate to configurations that are very different from that of which they were originally trained. Our fine-tuned version of the MACE-OMAT-MATPES foundation model is generated by adding 64 configurations from AIMD simulations (including the DFT-D3  dispersion correction) at 1000~K. From the original replay dataset of the MACE-OMAT-MATPES model \citep{Batatia2025,Kaplan2025_arxiv}, configurations containing only Mg and Cl atoms are kept as the  replay dataset to be used for the pre-training head. We use 64 DFT configurations as the input for the fine-tuning (FT) head. The multihead finetuning is performed using both of these datasets. 
The finetuning process can also be done by taking into consideration configurations at higher temperatures within the range of interest. We choose to do a minimal finetuning to check the improvement  on the MACE-OMAT-MATPES original  model and stick to configurations of a single temperature \citep{Batatia2025,Kaplan2025_arxiv}.

The  MACE architecture provides a machine-learning model of atomic interactions based on the training of a message-passing Neural Network (MPNN) \citep{Bronstein2021}, which is a graph neural network (GNN) embedded in 3D space. 
It unifies the Atomic Cluster Expansion (ACE) with equivariant message passing with high-body order, usually four-body order per layer. As a result, only two message-passing iterations are required  to achieve high accuracy. In addition, it uses tensor decomposition, allowing for 
efficient parallelization and scalability \citep{Batatia2022_mace_seminal,Batatia2025}. 
MACE  models, unlike the DeepMD model we generated for this work,  are equivariant by design, i.e. physical quantities transform under translations and rotations properly. The energy of the system remains invariant under rotations, while the forces transform directly as vectors under rotations.

\section{Results}


In the following paragraphs, we present the behavior of various physicochemical and structural properties of molten MgCl\textsubscript{2} in the 1000--1600~K temperature range. We use and compare  AIMD, the trained MLIP with DeepMD architecture, the MACE-MP-0b3 and MACE-OMAT-MATPES foundation models, including its fine-tuned version. We also compare our results  with experiments, whenever the experimental data is available.

\subsection{\label{subsec:Density}Density}

\begin{figure}
\centering{}\includegraphics[width=0.98\columnwidth]{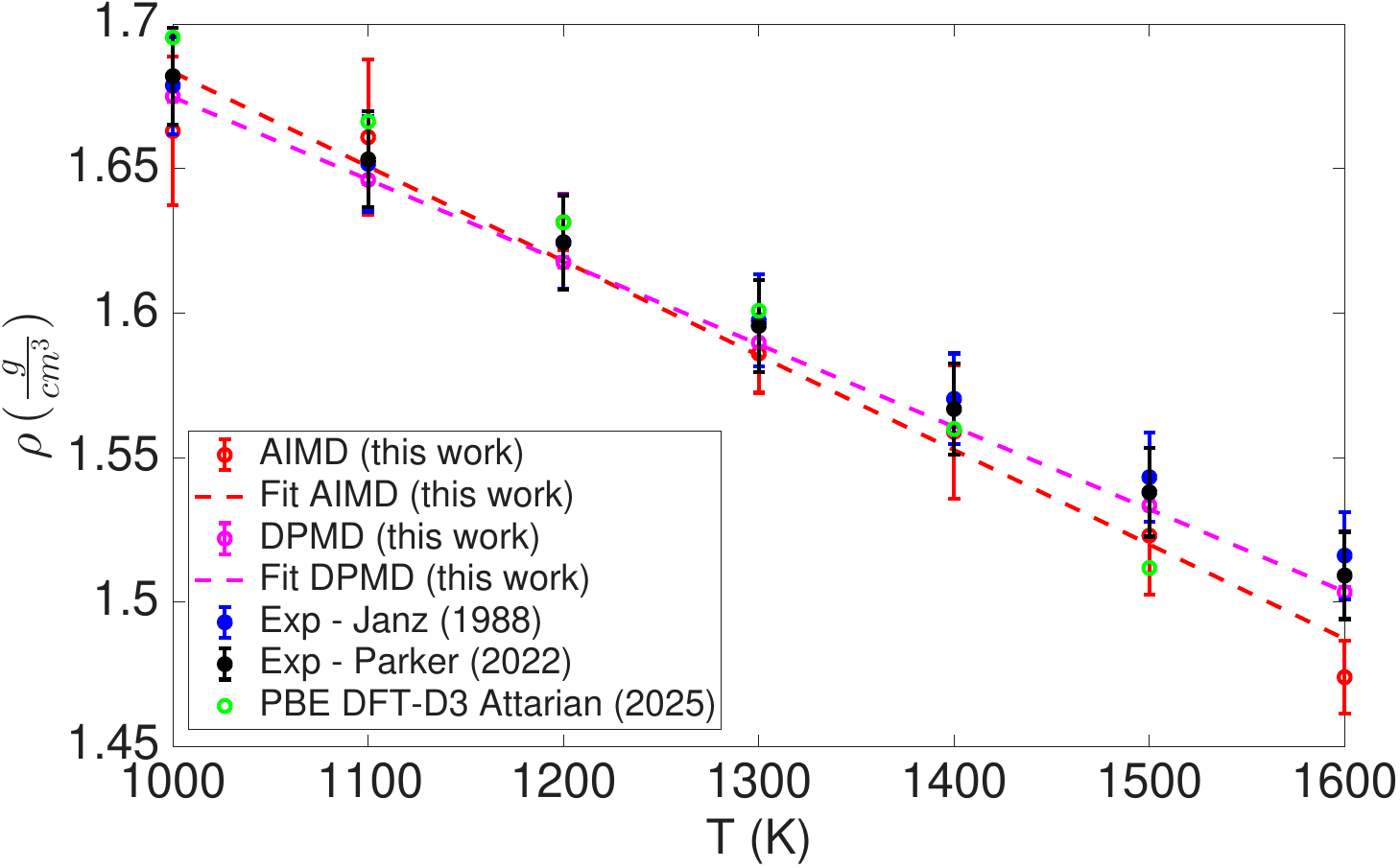}\caption{\label{fig:Density-vs.-temperature}Density vs. temperature for molten
MgCl\protect\textsubscript{2} in the 1000\protect\nobreakdash-1600$\,$K
range. We show the results of AIMD and DPMD simulations generated in this work and compare with two experimental works \citep{Janz1988,Parker2022} and a simulation work with another MLIP model \citep{ATTARIAN2025113409}.}
\end{figure}

In Figure \ref{fig:Density-vs.-temperature}, we show the temperature
dependence of molten MgCl\textsubscript{2} density. For AIMD simulations, density values go from 1.663~g/cm\textsuperscript{3} (1000~K) to 1.474~g/cm\textsuperscript{3} (1600~K),
decreasing upon increase of temperature, as expected.

The slope of the linear fit (see equation \ref{eq:dens_vs_T_AIMD}) is slightly higher (in absolute value) than the experimental slopes informed by Janz \citep{Janz1988} and Parker \citep{Parker2022}.
These are $\mathord{-}2.712~\times~10^{-4}~g/(cm^{3}~K)$
and $\mathord{-}2.88~\times~10^{-4}~g/(cm^{3}~K)$,
respectively. 
The percentage difference between our simulated values and the experimental data  of Janz increases with temperature reaching  a maximum absolute value of 2.78~\% at 1600~K.
While this difference can be considered significant, it should be taken into consideration that it is only at this temperature where no overlap
between simulated and experimental values is found. The error bars
obtained for the other temperatures have a significant overlap with
the experimental values. Given how difficult it is to obtain reliable experimental data at high temperatures, we think these values are in a reasonable agreement with it. From our calculations, we obtain a density-temperature relationship according to the following equation:

\begin{equation}
\rho_{_{{\rm AIMD}}}[\frac{g}{cm^{3}}]=2.011-3.269\times10{}^{-4}\,T\label{eq:dens_vs_T_AIMD}
\end{equation}

We  test  the quality of the our MLIP potential, by calculating the density of  molten MgCl\textsubscript{2} as a function of temperature, performing simulations in the NPT ensemble.
We use a sample of  4320~atoms in a cubic cell set to zero pressure. We use as  initial volumes of the MD cells those obtained as equilibrium
volumes by the Birch-Murnaghan method. Each run at a given temperature
is 200~ps long, the  first 100~ps are discarded as they are considered
part of the thermal equilibration of the system. The last 100~ps
are divided into 10~blocks of equal length in order to ``block-average''
the mean density and its uncertainty.
The fit for the density with the DPMD data is the following:

\begin{equation}
\rho_{DPMD}\,[\nicefrac{g}{cm^{3}}]=1.960-2.852\,\times\,10^{-4}\,T
\end{equation}

The slope of the fit is closer to both experimental results, showing  that the MLIP used in this
simulation is in good agreement with previous data and outperforms AIMD
results. 
The density values go from 1.675~g/cm\textsuperscript{3} at 1000~K to 1.503~g/cm\textsuperscript{3} at 1600~K, which are higher and closer to the experimental values of Janz et. al \citep{Janz1988}
and Parker et. al \citep{Parker2022}. Density values calculated using DFT by Attarian et al. are also shown for comparison \citep{ATTARIAN2025113409}.






\subsection{\label{subsec:Radial-distribution-function}Radial distribution function
(RDF)}

\begin{figure}
\centering{}\includegraphics[width=0.98\columnwidth]{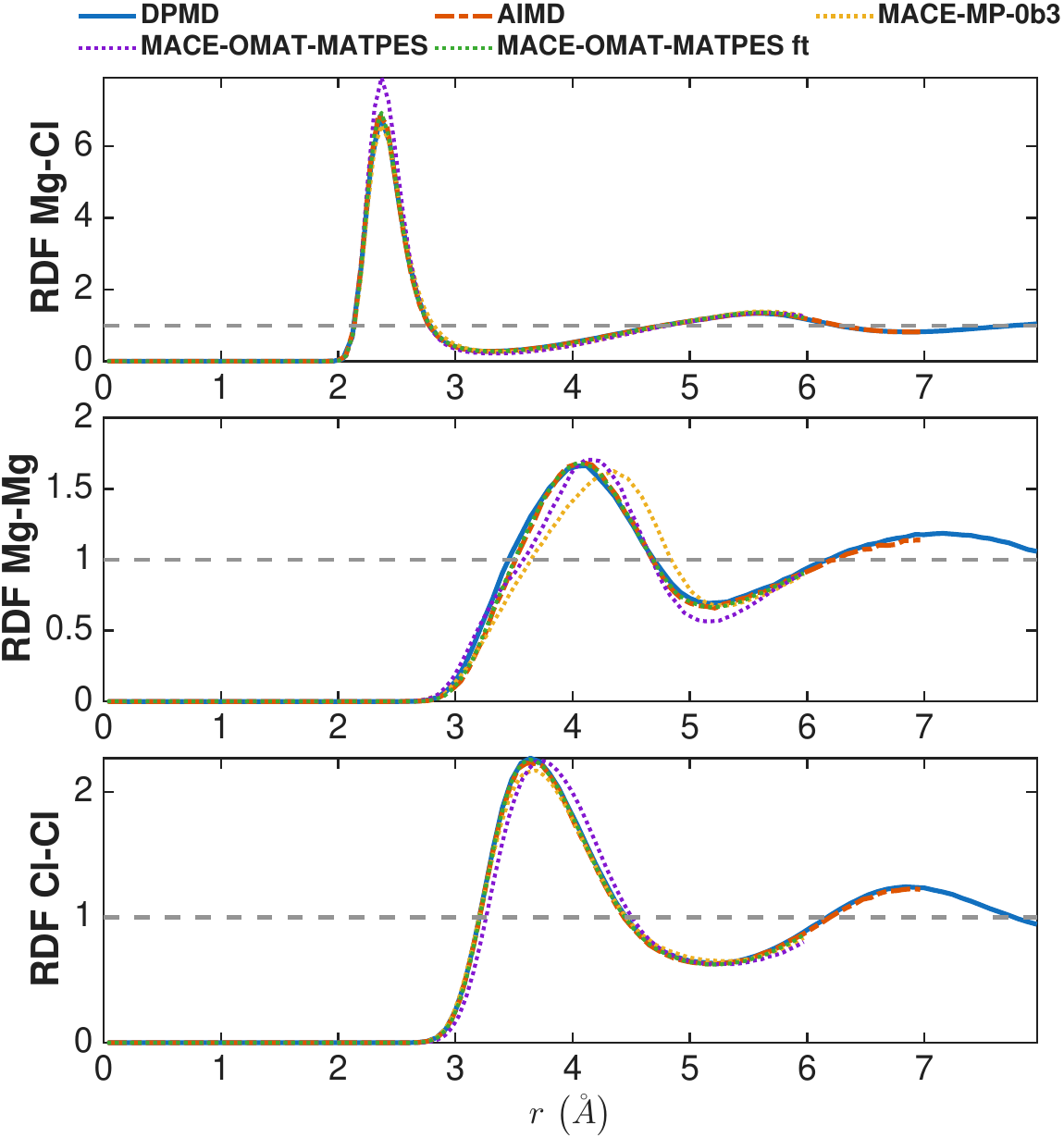}\caption{\label{fig:MgCl2-RDFs-1000K} RDFs for molten MgCl\protect\textsubscript{2} at 1000~K obtained by AIMD, DPMD, MACE-MP-0b3, MACE-OMAT-MATPES and a fine-tuned MACE-OMAT-MATPES model. The pairs Mg-Cl, Mg-Mg and Cl-Cl are shown in the upper, center and lower panels, respectively.}
\end{figure}

To characterize the structural properties of this molten salt, we 
analyze the radial distribution functions (RDF) corresponding to the Mg-Cl, Mg-Mg and Cl-Cl ion pairs.
From these data we extract structural parameters,  including
the average interionic distances  and the coordination numbers. In Figure
\ref{fig:MgCl2-RDFs-1000K} we show  the RDFs  at 1000~K, calculated by AIMD, DPMD, MACE-MP-0b3, MACE-OMAT-MATPES and our fine-tuned version of MACE-OMAT-MATPES.

AIMD and DPMD produce closely matching RDFs for all the ionic pairs
 up to about 7\,\AA. This means that the structures predicted by
AIMD and DPMD are essentially the same up to the second coordination shell.
The positions of the first maximum of the Mg-Cl RDF go from 2.370~Å (1000~K) to 2.352~Å (1600~K) in the studied temperature range.
In the case of Mg-Mg and Cl-Cl distances, the positions of the first maximum are in the ranges 4.052\,-\,4.153~Å (1000-1600~K) and 3.626\,-\,3.663~Å (1000-1600~K), respectively.
The Mg-Cl average distance is slightly lower than the value (2.42 \,\textpm \,0.03)\,Å obtained by Biggin et al \citep{Biggin1984}.
For higher temperatures, the general features of the RDFs are very similar to this case and the results are presented in Figures S1 and S2 of the Supplementary Material.

The RDFs obtained with the out-of-the-box MACE-MP-0b3 foundation model for the pairs  Mg-Cl and Cl-Cl are in very good
agreement with AIMD and DPMD results (upper and lower panels in Fig. \ref{fig:MgCl2-RDFs-1000K}).
  However, there is a discrepancy in the location of the first peak of the Mg-Mg pair (dashed orange curve, center panel in Fig. \ref{fig:MgCl2-RDFs-1000K}), which is located at larger distance as compared with the AIMD and DPMD cases. 

The fact that the  out-of-the-box model gives such good results is quite remarkable. Therefore, we decided to try  it even without
employing the DFT-D3 dispersion correction, to test its accuracy
on other structural and dynamical properties of this molten salt.

Comparing our results with those of Liang et al. \citep{Liang2020},
we observe that the Mg-Cl and the Cl-Cl average distances are very similar,  but there is a discrepancy in the case of the Mg-Mg average distance. 
The value reported by them is 3.871~Å, which is between 4.7$\,$\% and 7.3$\,$\% below the range of values we find.
Since they also perform AIMD simulations, we attribute this difference to the use of the DFT-D3 correction in our calculations to account for dispersion forces, whereas DFT-D2 is used in their work. Both our RDFs and the ones obtained
by Liang et al. \citep{Liang2020} differ from the experimental
RDFs presented by Wilson and Madden \citep{Wilson1993}.
There, it can be clearly observed that Mg-Mg and Cl-Cl RDFs have their
first maximum at the same position of about 3.64~Å. The RDFs of Mg-Mg
and Cl-Cl pairs are shown in Figure S1 of the Supplementary Material.

\begin{figure}
\centering{}\includegraphics[width=0.98\columnwidth]{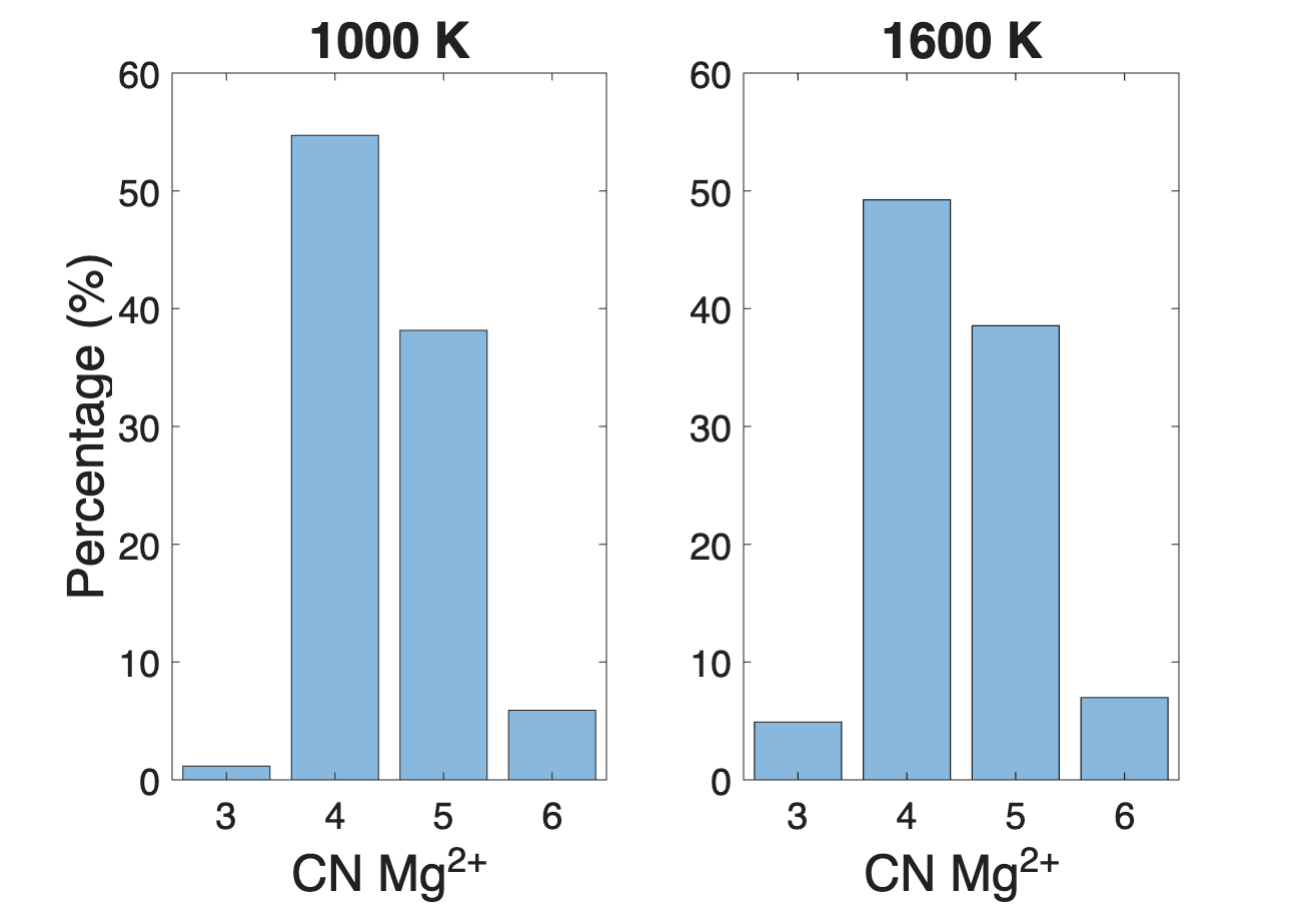}
\caption{\label{fig:Coordination-number-distribution}
Coordination number distribution of Cl ions around Mg ions at 1000$\,$K and 1600\,K, obtained from the RDFs calculated for the AIMD simulations.}
\end{figure}

By integration of the partial RDFs according to the expression:

\begin{equation}
N_{\alpha\beta}=\int_0^{r_{min}}4\pi r^{2}\rho_{\beta}g_{\alpha\beta}(r)\,dr~,
\end{equation}

\noindent we get the coordination numbers of the first coordination shell of
Cl\textsuperscript{-} ions around Mg\textsuperscript{2+} ions.
There, \(r_{min}\) is the position of the local minimum between the first and the second peaks of the RDF. We obtain values going from 4.51 (1000~K) to 4.48 (1600~K).
For Mg–Mg and Cl–Cl pairs, the corresponding ranges are 5.08–5.65 (1000-1600~K) and 11.38–12.48 (1000-1600~K), respectively. Biggin et al. \citep{Biggin1984} measure the following coordination numbers experimentally: $4.3\pm0.3$ (Mg-Cl), $5\pm1$ (Mg-Mg) and $12\pm1$ (Cl-Cl), which are in good agreement with our results. Our calculations are also in good agreement
with previous results by Liang et al. \citep{Liang2020}.  For Mg\textsuperscript{2+}, Wilson and Madden obtain a coordination number of 5.45, which differs considerably from the other simulation and experimental results. We also calculate the distribution of the coordination numbers across the system for the RDFs obtained by AIMD simulations. The different coordination numbers are presented in Figure \ref{fig:Coordination-number-distribution}
for the two extreme temperatures, 1000~K and 1600~K, of the studied range.

Four-fold ($\sim$\,55\,\% for 1000~K and $\sim$\,49\,\% for
1600~K) and five-fold ($\sim$\,38\,\% for both temperatures, 1000~K and 1600~K)
coordination numbers are predominant in the system. However, coordination numbers as low as 3 ($\sim$\,1\,\% for 1000\,K
and $\sim$\,5\,\% for 1600~K) and as high as 6 ($\sim$\,6\,\%
for 1000~K and $\sim$\,7\,\% for 1600~K) are also observed. As temperature
rises, the distribution widens and flattens, keeping the same overall
structure. However, at 1600~K it is possible to observe that some two-fold
($\sim$\,0.01\,\%) and eight-fold ($\sim$\,0.004\,\%) cases are present.
The distribution of four-fold and five-fold coordination numbers is
different from the one shown in Liang et al. \citep{Liang2020}.
They found that five-fold coordination is the most abundant at 1000~K,
1200~K and 1400~K, closely followed by four-fold coordination.
Six-fold coordination is more abundant in their AIMD simulations ($\sim$\,10\,\% or more) and three-fold coordination is less abundant than in our study. In both works, the distribution flattens and widens, as expected,  as temperature increases.

\subsection{\label{subsec:Angular-distribution-function}Angular distribution
function}

\begin{figure}
\centering{}\includegraphics[width=0.98\columnwidth]{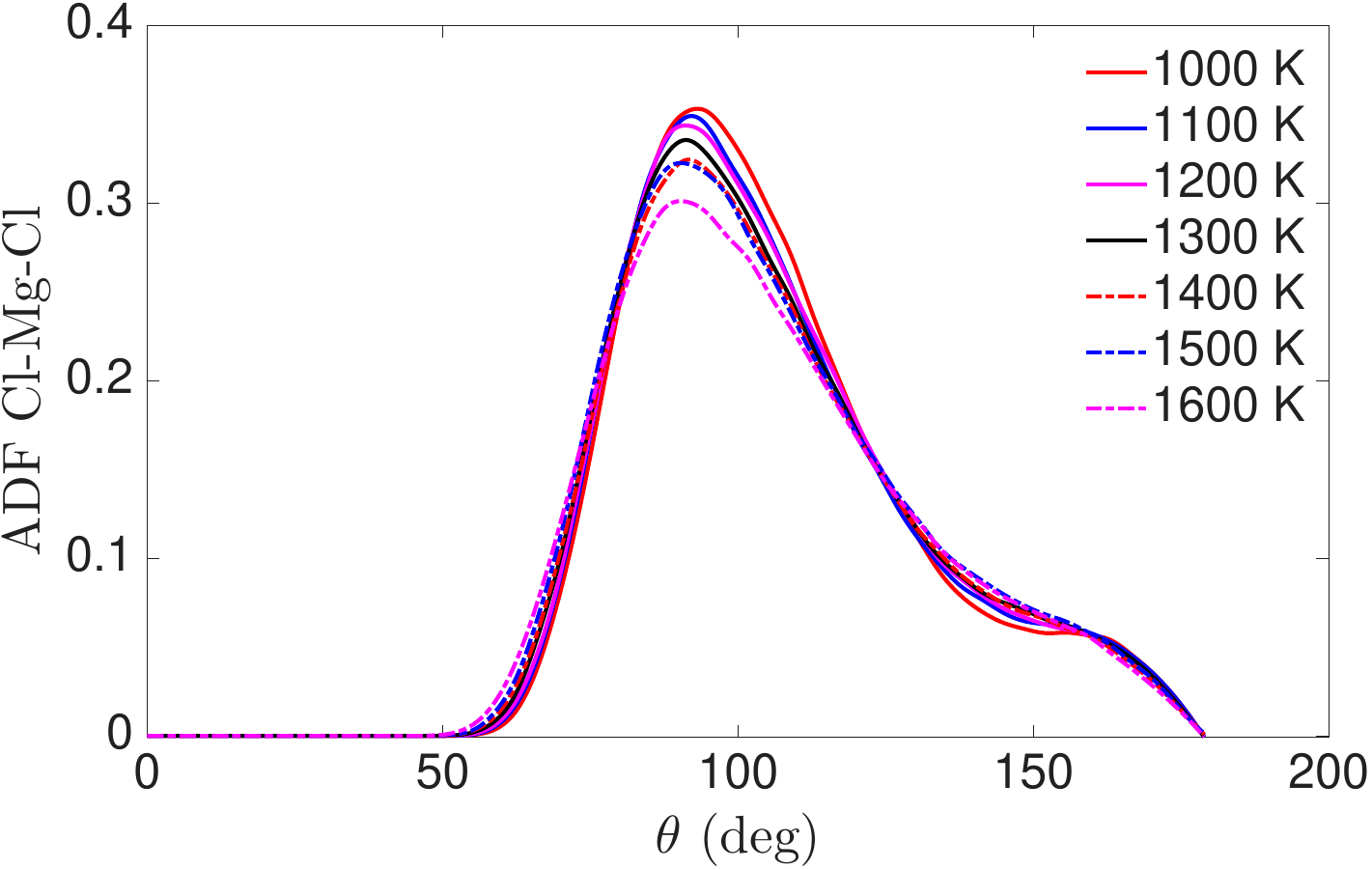}\caption{\label{fig:Angular-distribution-function}Angular distribution function
of Cl-Mg-Cl triplets in molten MgCl\protect\textsubscript{2} in the
1000\protect\nobreakdash-1600$\,$K range as obtained from AIMD simulations.}
\end{figure}

The angle formed by the Cl-Mg-Cl triplet, where the Cl\textsuperscript{-} ions are in the first coordination shell of Mg\textsuperscript{2+} ions, is an important measure of the degree of structural deformation in molten MgCl\textsubscript{2} at high temperatures. We show the angular
distribution function for the Cl-Mg-Cl triplet in Figure \ref{fig:Angular-distribution-function}. A marked peak is observed at an angle of $\approx93^{\circ}$ for all the temperatures. The position of this peak is quite insensitive to temperature, given the angular resolution of the ADFs. This behavior is in correspondence with the insensitivity to temperature observed for the RDFs. There is also a secondary peak at
the 165$^{\circ}$ angle, with a clear shoulder at lower temperatures.

As in the case of the Mg-Cl RDF, there is no significant change in
the position of the maximum of the ADF with temperature. The maximum
at $\approx 93^{\circ}$ shows that many of the molten MgCl\textsubscript{n}
(\(n = 3\) to \(6\)) complexes have triplet angles that are similar to
the those characteristic of the octahedral coordination in solid MgCl\textsubscript{2}, albeit a significantly spread out due to thermal effects. The main type of structure is now tetrahedral. The
shoulder at $\approx165^{\circ}$ corresponds to the appearance of greatly distorted
MgCl\textsubscript{4} and MgCl\textsubscript{5} complexes. The shoulder
gradually smooths out at high temperatures, although it never vanishes
completely. We recall that the triplet angles are either 90$^{\circ}$ or 180$^{\circ}$ for solid MgCl\textsubscript{2} salt.

\subsection{\label{subsec:Thermal-conductivity}Thermal conductivity}

Thermal conductivity is a key property for the technological applications of molten MgCl\textsubscript{2} and it has been determined with great uncertainty in the literature.
Very different experimental  results are frequenty found, due to the difficulties in the experimental setup and heat losses induced both by convection and thermal radiation \citep{Gheribi_2026}. Experiments are usually  labeled as ``reliable'' or ``unreliable'', as can be observed in Fig. 1 of Ref. \citep{Gheribi2022} for NaCl.   For MgCl\textsubscript{2} the situation is even worse, with many unreliable  experiments and others done for a small range of temperatures. See for example Fig. 3 of Ref.  \citep{Gheribi2022}. 

We calculate the thermal conductivity  as a function of temperature at 1000$\,$K, 1200$\,$K, 1400$\,$K and 1600$\,$K with out-of-equilibrium MD simulations using our DP model.

\begin{figure}
\centering{}
\includegraphics[width=0.98\columnwidth]{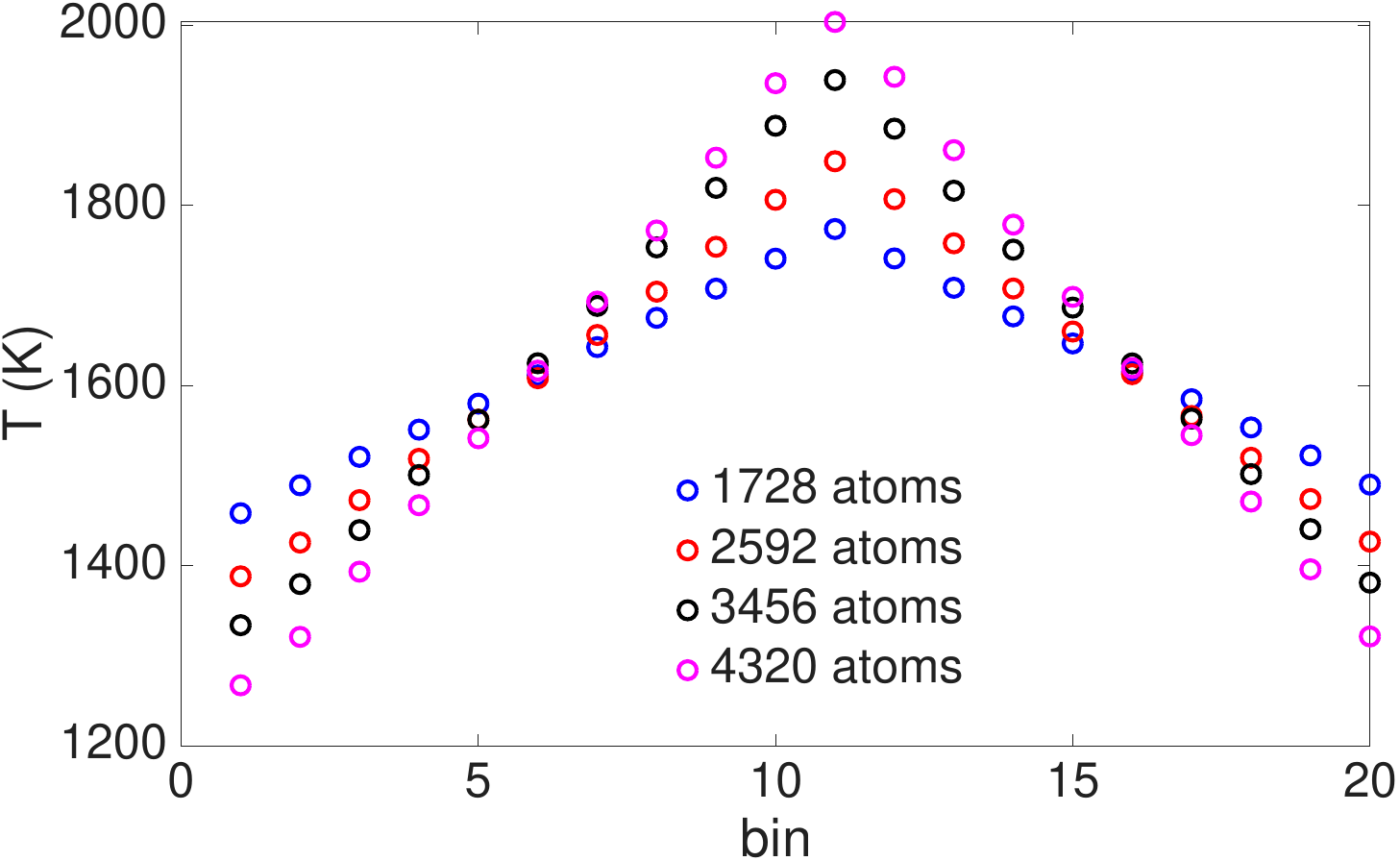}\caption{\label{fig:Temperature-profile}
Temperature profiles in for molten MgCl\protect\textsubscript{2} at 1600$\,$K generated by the Müller-Plathe method as obtained from DPMD simulations. The increasing  number of atoms correspond to samples of progressively larger lengths in the direction of the heat flux $L_z$. Each bin corresponds to a division of the simulation cell into 20 ``slabs''. }
\end{figure}

A temperature profile is generated along the $z$ direction by the method of Müller-Plathe, as explained in Section \ref{subsec:Thermal-conductivity-1}. 
The  mean temperature profiles for systems of different lengths $L_z$ in the direction of the heat flux  are shown in Figure \ref{fig:Temperature-profile}. The linear  profiles between the extremes of the sample (cold zone) and the center (hot zone) are evident, arising from the stationary state of the velocity exchange procedure.
The transient states last a high number of time steps (we discard the first 2~ns for the thermal conductivity calculation) in which the temperature profiles progressively build up. After this transient and when the profiles stop varying in time, the average temperatures in the  bins are calculated.

The slopes of each profile, provide a value of the thermal conductivity
for  different  system sizes and can be calculated according to Equation \ref{eq:fourier_heat} from Section \ref{subsec:Deep-Potential-Molecular-1}.
We vary the sample length to obtain a  thermal conductivity value for each box length, which are then extrapolated to the limiting case of vanishing thermal gradient, which is called the size-independent thermal conductivity $\lambda_{SI}$. 
The technical details of the procedure are explained in Section II of Supplementary Material, along with a study of the sensitivity of the thermal conductivity values with the ion velocity exchange rate.
We find  that it is quite independent  on the choice of the exchange rate  (see Fig.~S5 and the related text in Supplementary Material). 



In Figure \ref{fig:TCsi_vs_rho_MP200} we show the thermal conductivity \(\lambda\) as a function of the system density. An increase of the thermal conductivity is observed upon density increase, with a saturation at higher densities, which correspond to lower temperatures (indicated below the symbols in Fig. \ref{fig:TCsi_vs_rho_MP200}). 

\begin{figure}[h]
    \centering
    \includegraphics[width=0.98\columnwidth]{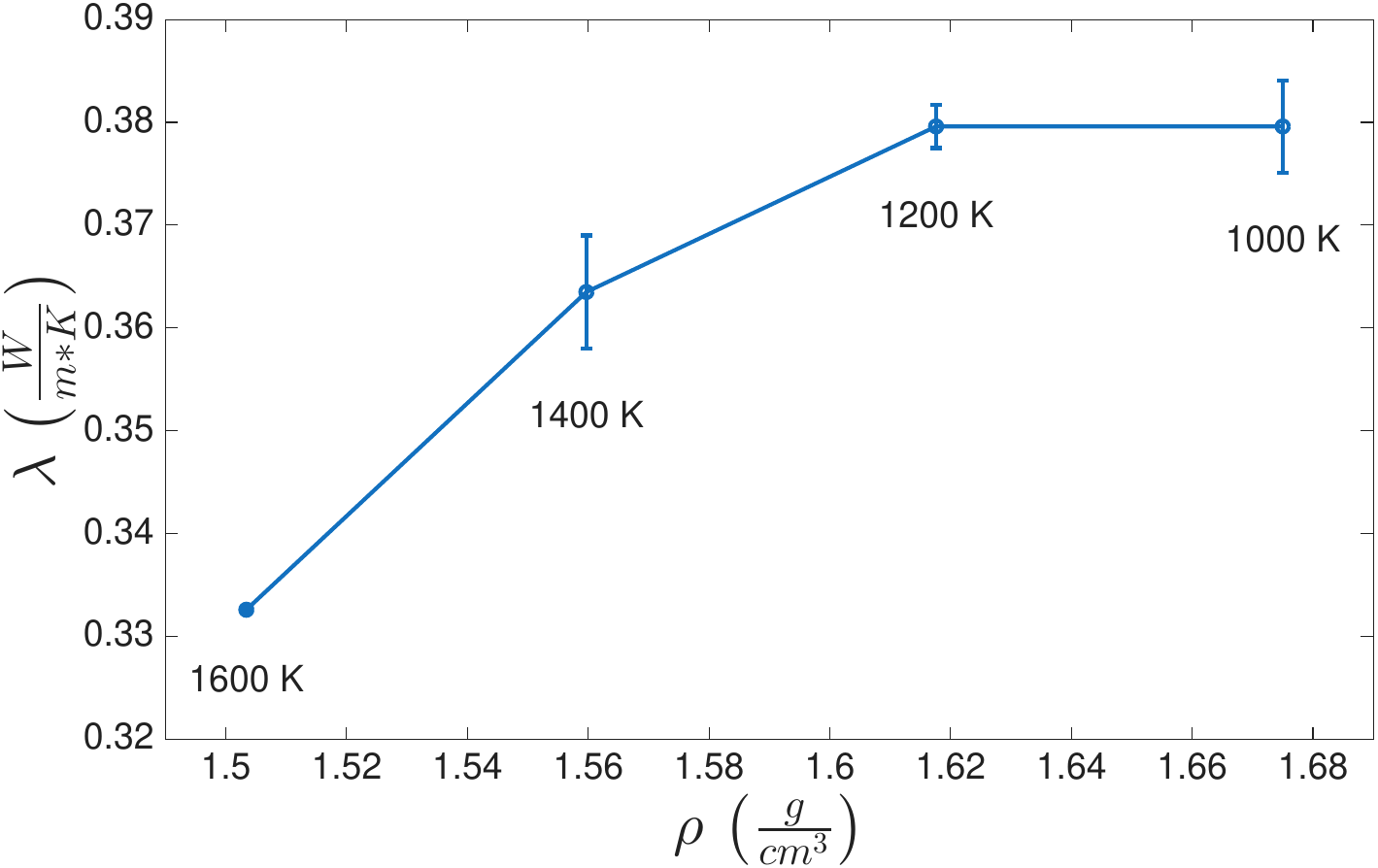}
    \caption{Thermal conductivity of molten MgCl\textsubscript{2} vs.  density as given by simulations of our DP model. The temperature values are also indicated. }
    \label{fig:TCsi_vs_rho_MP200}
\end{figure}

\begin{figure}[h]
    \centering
    \includegraphics[width=0.98\linewidth]{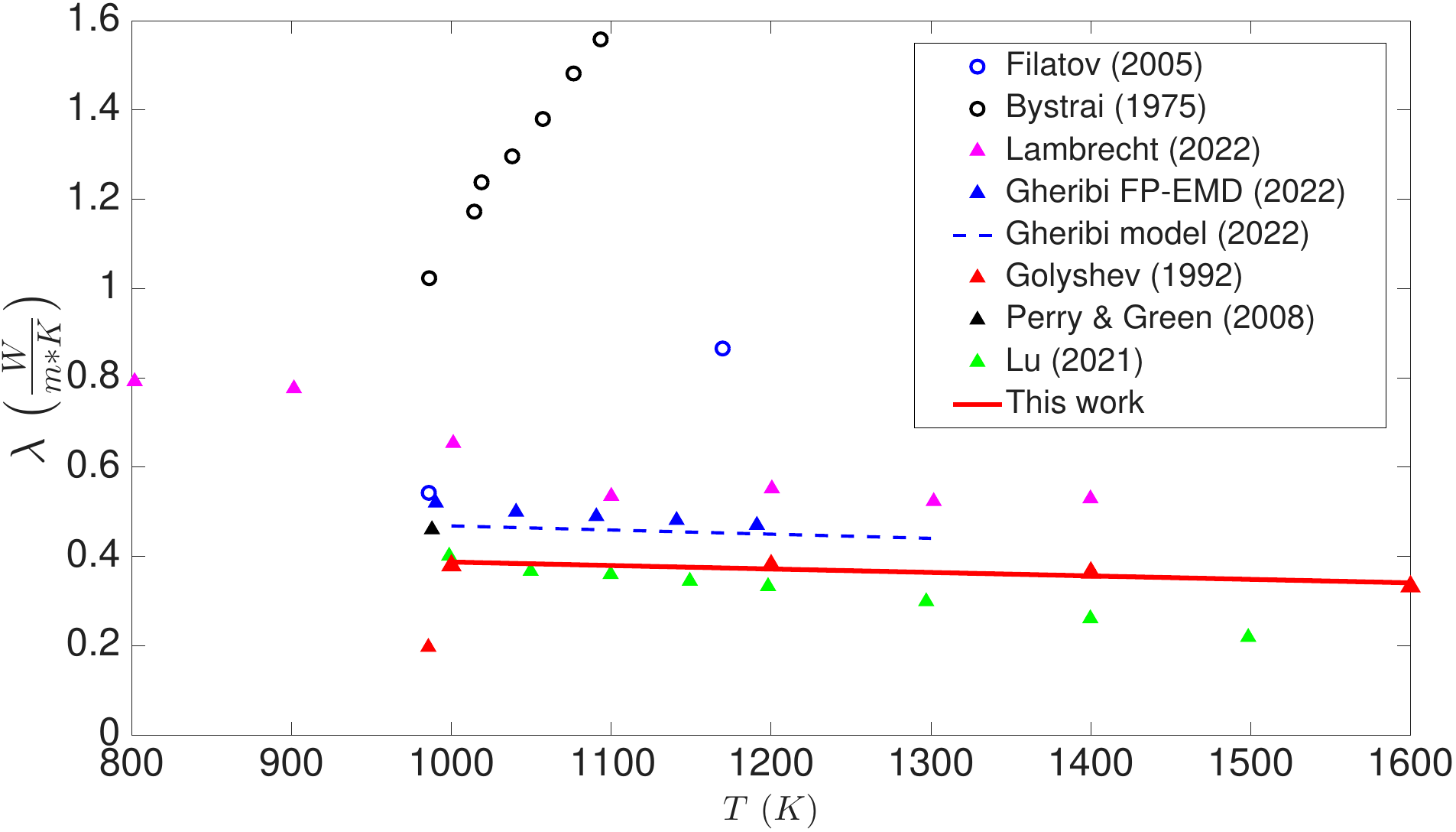}
    \caption{Thermal conductivity of molten MgCl\textsubscript{2} vs temperature. In red full circles our data from DPMD simulations is shown, with its linear fit (red line). A comprehensive set of simulation (filled triangles) and experimental results (hollow circles) is also shown for comparison, see the details in the text.}
    \label{fig:TCsi_vs_T_MP200}
\end{figure}

We show in Fig. \ref{fig:TCsi_vs_T_MP200} the  thermal conductivity of molten MgCl\textsubscript{2} vs. temperature as obtained from  our DPMD simulations. Data from previous experimental and theoretical results is also shown.

Gheribi et al. \citep{Gheribi2014, Gheribi2022} developed a model
by combining a previous result establishing
 that the thermal conductivity of molten salts 
 at constant volume $V$ does not depend on temperature
 ( \( \partial(\ln \lambda)/\partial T )_\rho =0)   \) ) and kinetic theory. 
They condense their  model in the Eq.:

\begin{equation}
\lambda_{\rm{Gheribi}}\,[\nicefrac{W}{(m*K)}]=0.5614-9.3\,\times\,10^{-5}\,T
\label{eq:25}
\end{equation}

\noindent see Eq. 13  and close paragraphs in Ref. \citep{Gheribi2022}. The model is shown in dashed blue line in Fig. \ref{fig:TCsi_vs_T_MP200} and also the results from molecular-dynamics simulations using a PIM interaction potential  from the same reference (blue triangles). 

We show our results in red circles, the red line corresponding to  the  linear fit  of the data, with Equation:


\begin{equation}
\lambda^{\rm DPMD}[\nicefrac{W}{(m*K)}]=0.466-7.8\,\times\,10^{-5}\,T
\end{equation}

Regarding simulation data, we also show the simulation results  by Lu et al. (green triangles),  who performed molecular-dynamics simulations of another PIM model \citep{Lu_2021b}. Our results agree very well at $T=1000~K$ but  the  (negative) slope of their thermal conductivity is much higher than ours. 
This is also true for the simulations of Gheribi et al., who  present systematic higher values of thermal conductivity and also a higher slope in the fall of \( \lambda \) with temperature (blue triangles). Simulations with PIM potentials are generally considered to predict thermal conductivity for molten salts compounds  with an accuracy in  the range  of 20\%, as compared with reliable experimental data \citep{Gheribi2022}. Our  data lie inside this confidence interval,  as compared with data from Lu et al. \citep{Lu_2021b} for the whole range of studied temperatures.
%
They report a slope of $2.59\times 10^{-4} \,Wm^{-1}K^{-2}$ which is significantly higher than the theoretical value of $9.3\times 10^{-5} \,Wm^{-1}K^{-2}$ (see Eq. \ref{eq:25}) from their thermodynamic model. We obtain a value for the slope of $\partial \lambda^{ \rm DPMD}/\partial T = 7.8\times 10 ^{-5}\,Wm^{-1}K^{-2}$, which is very close to the theoretical model of Eq. \ref{eq:25} \citep{Gheribi2022}.
From a comparative study of the values of thermal conductivities  among  alkali  earth fluoride salts above melting temperature, Gheribi et al. \citep{Gheribi2022}  mention an expected value of $\lambda
\simeq0.4\,W/mK$ close to the  melting temperature and above  $0.2\,W/mK$. This is consistent with our result of $\lambda (T=1000~K)=0.38\,W/mK$.
Regarding the experiments, our value of $\lambda(T=1000~K)$ lies in between the newer experimental works by Lambrecht et al. \citep{LAMBRECHT2022123273} (cyan circles), Perry and Green (black triangle) \citep{Green2019Perry} and Golyshev (red triangle) \citep{Golyshev1992}. Perry and Green's result are often reported as reliable, but unfortunately they report only one  value close to the melting temperature, without a dependence of $\lambda$ with temperature. 
The experimental results by Bystrai and Desyatnik (black circles in Fig. \ref{fig:TCsi_vs_T_MP200}) \citep{bystrai1975thermal} and Filatov et al. (blue circles) \citep{Filatov} are shown for completeness, but they are considered unreliable, because the increase of thermal conductivity with temperature is unphysical and a possible indicator of heat losses in the experiments \citep{Gheribi2022}.

To make a conclusive evaluation  of the results more experimental data is needed, but the overall agreement of our data with other simulations and experimental results indicates that the MLIPs simulations are a promising route to obtain reliable data for molten salts, making possible the use of 
non-equilibrium simulations to compute transport properties.  






\subsection{\label{subsec:Heat-capacity-1}Heat capacity}

\begin{figure}
\begin{centering}
\includegraphics[width=0.98\columnwidth]{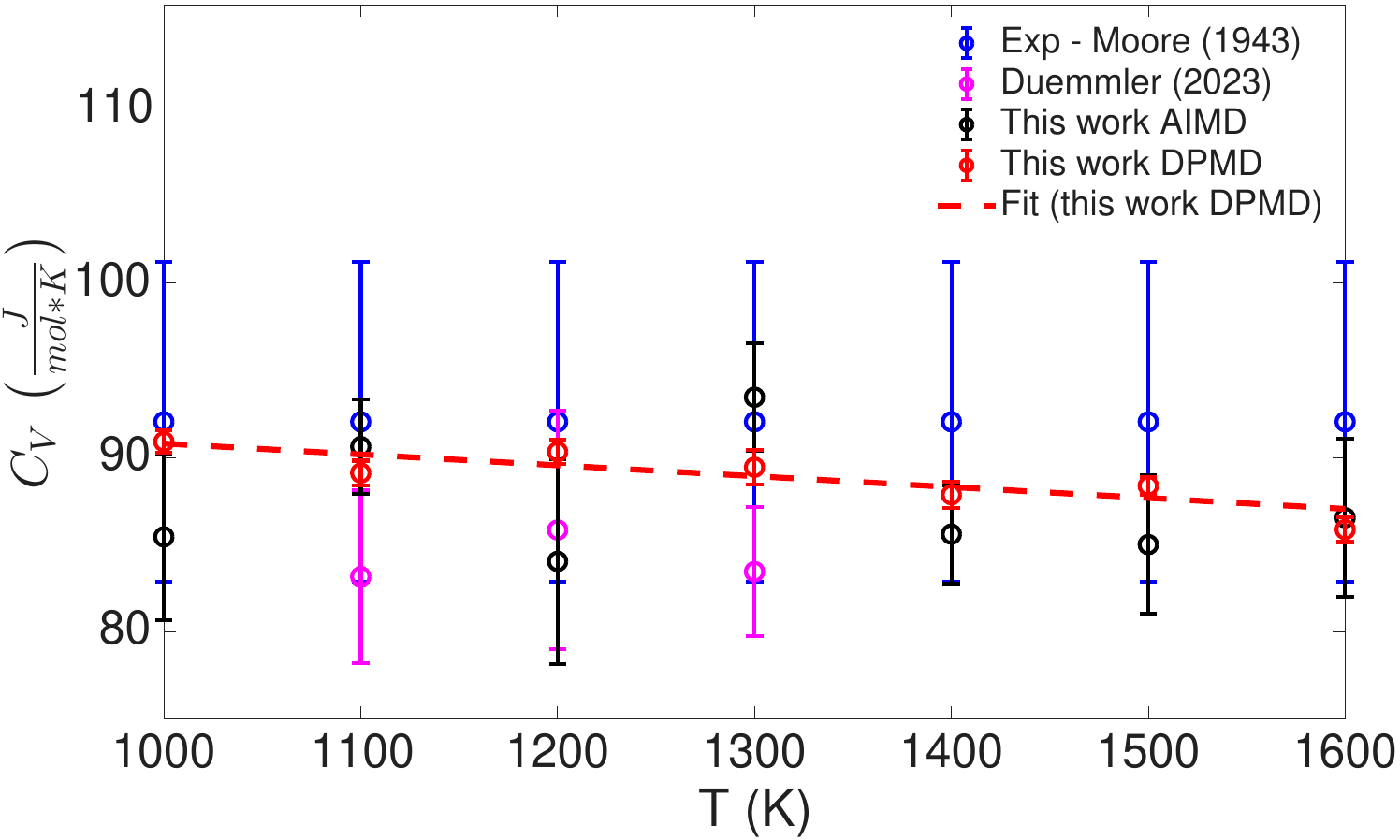}
\par
\end{centering}
\caption{\label{fig:Heat-capacity-of}Heat capacity of molten MgCl\protect\textsubscript{2}
in the 1000-1600~K range, obtained with simulations using the DP model. Our AIMD results and other data from the literature are also shown.}
\end{figure}

With the largest sample used for the thermal conductivity calculations, we calculate the heat capacity at constant volume employing the PES generated in Section \ref{subsec:Deep-Potential-generation-1}.
The simulation time interval is divided into 4 blocks to calculate mean
values and uncertainties. The results obtained from DPMD are presented in Figure \ref{fig:Heat-capacity-of} (red symbols and fit) and compared with experimental results by Moore et al. \citep{Moore1943} (blue symbols) and simulation data by Duemmler et al. \citep{Duemmler2023} (cyan symbols).
Our simulated data is in very good agreement and lies between the other two sets of results. 
For  the sake of comparison, the values obtained from AIMD are also presented in black symbols.  The higher errors are evident, as expected from much shorter trajectories of a smaller sample. 

The linear fit for the C\textsubscript{V} values obtained by DPMD
is as follows:

\begin{equation}
C_{V}[\nicefrac{J}{(mol\,K)}]=97.05-6.25\times10^{-3}\,T
\label{eq:Cv_fit_500fs}
\end{equation}

The negative slope of the fit, indicates a slight decrease of
the heat capacity upon increase of temperature. It is clearly observed  that
the use of a larger system entails an increase in the precision of the values obtained for the heat capacity.
The values above are obtained with a damping factor of 500~fs.

The slope in Equation \ref{eq:Cv_fit_500fs} has a relative  error of  $\sim\,25\%$, implying a decreasing  C\textsubscript{V} as a function of  temperature in our DPMD simulations. C\textsubscript{V} decreases $\approx5.4\%$ (compared to its value at 1000~K) in the studied temperature range ($C_{V}(T=1000~K) \approx 90.8\,J/(mol\,K)$ and $C_{V}(T=1600~K) \approx 85.86\,J/(mol\,K)$). This suggests a reduction  of this salt's ability to storage thermal energy at the higher temperatures in the studied range (from $\approx1300~K$ onwards).

\subsection{\label{subsec:Viscosity-1}Viscosity}

We obtain the viscosity as a function of temperature  from equilibrium MD simulations using our DP model, the MACE-MP-0b3, the MACE-OMAT-MATPES and also a fine-tuned version of that uMLIP model. We calculate the autocorrelation function (SACF) of the non-diagonal elements of the pressure tensor as explained in subsection \ref{subsec:Viscosity}.

 We show a plot of the average SACF as a function of the correlation time for molten MgCl\textsubscript{2} at 1000~K in Figure~S5  of  the Supplementary Material. We note  that it decays rapidly, with a first characteristic time $\tau_{1}\simeq21\,fs$. Then, it continues its decay with a second characteristic time $\tau_{2}\simeq0.82\,ps$ in this case. This autocorrelation function
can be fitted satisfactorily with a double exponential function:

\begin{equation}
\rm{SACF}(t)=a\,\exp(-\frac{t}{\tau_{1}})+b\,\exp(-\frac{t}{\tau_{2}})
\end{equation}

Table \ref{tab:Characteristic-decay-times} shows the mean values of these characteristic times as a function of temperature calculated over ten  statistically independent simulations. $\tau_{1}$ shows a behavior that is quite independent of temperature, with a value $\approx21\,{\rm fs}$ over the whole temperature range. This indicates that the fast mechanisms that decorrelate the pressure tensor are not influenced significantly by temperature.
On the other hand, we observe that $\tau_{2}$ clearly diminishes with temperature. This suggests that the slow response of the system is related to the diffusion and the momentum transport mechanisms within the salt. They are influenced by temperature and are expected to be faster at higher temperatures. 

\begin{table}

\begin{centering}
\begin{tabular}{|c|c|c|}
\hline 
Temperature (K) & $\tau_{1}(fs)$ & $\tau_{2}(ps)$\tabularnewline
\hline 
\hline 
1000 & 20.8(1) & 0.71(3)\tabularnewline
\hline 
1100 & 20.8(1) & 0.51(1)\tabularnewline
\hline 
1200 & 20.9(1) & 0.44(2)\tabularnewline
\hline 
1300 & 20.8(1) & 0.38(1)\tabularnewline
\hline 
1400 & 21.0(1) & 0.35(1)\tabularnewline
\hline 
1500 & 21.1(1) & 0.31(1)\tabularnewline
\hline 
1600 & 21.1(1) & 0.29(1)\tabularnewline
\hline 
\end{tabular}\caption{\label{tab:Characteristic-decay-times}
Mean characteristic decay times $\tau_{1}$ and $\tau_{2}$ of the
pressure tensor non-diagonal components SACF as a function of temperature for molten MgCl\protect\textsubscript{2} in the 1000-1600~K range. Averages are calculated over ten thermalized and statistically independent simulations.}
\par\end{centering}
\end{table}

Integrating the cumulative SACF, we obtain a converging  plot of viscosity vs. time (number of autocorrelation windows averaged) from which we can verify if the total simulation time is long enough such that the  viscosity value is converged.
An example of this plot is shown in Figure S6 of the Supplementary Material.
There, it can be observed  that the viscosity fluctuates strongly for shorter times, but converges to a final value beyond $\approx 0.5\,{\rm ns}$. 
The experimental value by Tørklep and Øye \citep{Toerklep_1982} is also shown in Figure S6. For the different statistically independent initial configurations, convergence can be slower or faster and approaching from above or below the experimental value. By averaging the converged values of $\eta$ from ten independent runs at each temperature, we determine a final mean viscosity value and its standard deviation. We refer the reader to Section III of Supplementary Material for more technical details on the calculation of viscosity.

\begin{figure}

\begin{centering}
\includegraphics[width=0.98\columnwidth]{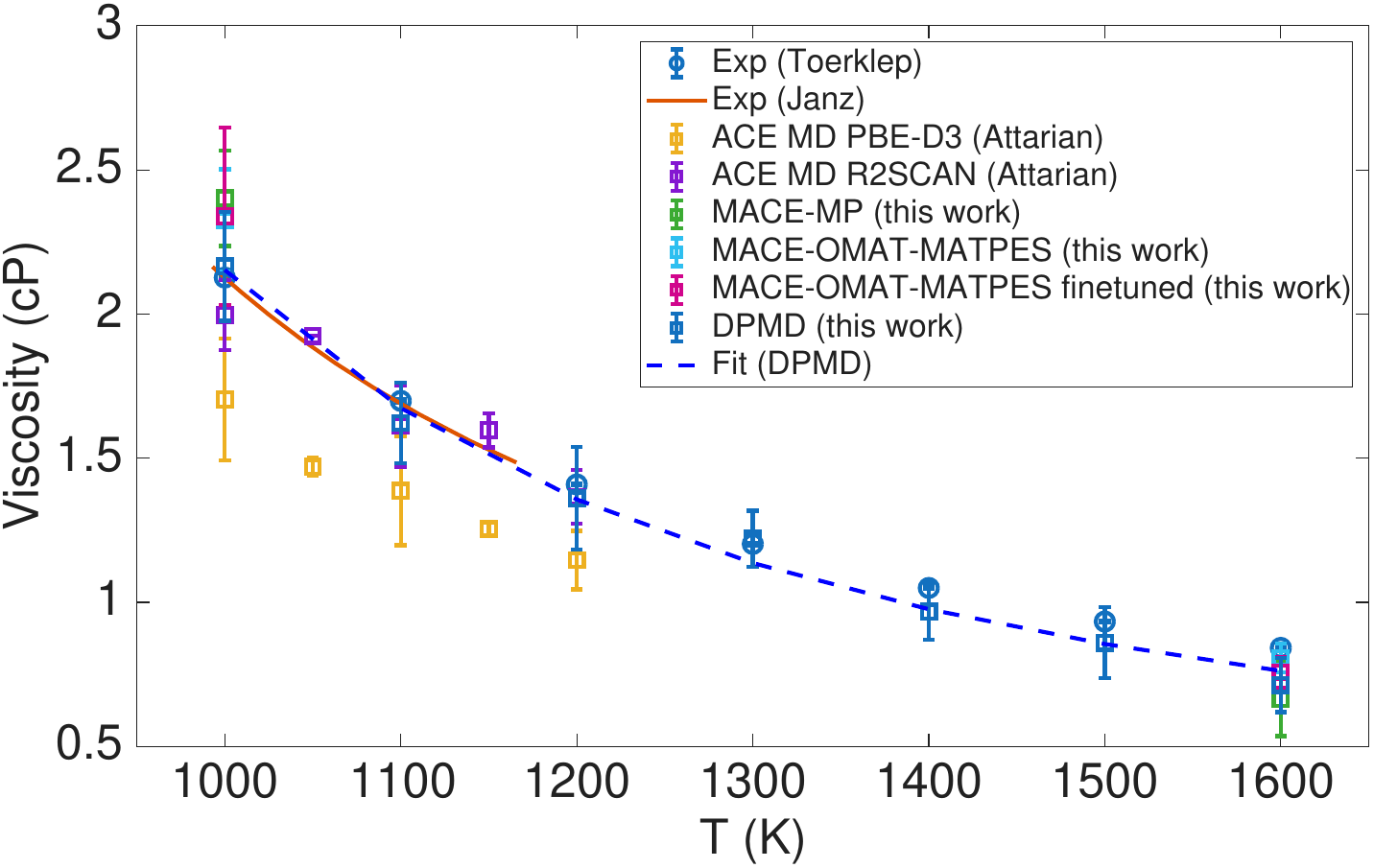}\caption{\label{fig:Viscosity-vs.-temperature}Viscosity vs. temperature for
molten MgCl\protect\textsubscript{2} in the 1000--1600~K range. Simulations with various models (hollow squares) and experimental results (hollow circles) are compared.}
\par\end{centering}
\end{figure}

In Figure \ref{fig:Viscosity-vs.-temperature} we show the viscosity values obtained for our DP model (blue triangles), the MACE-OMAT-MATPES out-of-the-box model (cyan circles), the fine-tuned version of the model (violet triangles) and various experimental and simulations works that we will describe in the next paragraphs.
The exponential fit of the DPMD data (dashed blue line in Figure \ref{fig:Viscosity-vs.-temperature}) presents the following equation:

\begin{equation}
\eta\,[cP]=0.1351\,\exp(\nicefrac{2768}{T})
\end{equation}

In the  experimental work by Tørklep and Øye \citep{Toerklep_1982}, they fit the experimental data to the following expression

\begin{equation}
\eta_{Toerklep}[cP]=0.18\,\exp(\nicefrac{2470.1}{T}).
\end{equation}

While the experimental work by  Janz \citep{Janz1988} presents  equation, which is shown in red line in Figure  \ref{fig:Viscosity-vs.-temperature}:

\begin{equation}
\eta_{Janz}[cP]=0.1794\,\exp(\nicefrac{2472.6}{T})
\end{equation}

 We observe  an excellent agreement between  our results from the DP model and  the fits of the experimental works by  Tørklep \citep{Toerklep_1982}  and Øye, and Janz \citep{Janz1988}. 
 This agreement is remarkable because it allows a comparison for the whole  range of experimental data, showing that the simulations are able to be an effective predicting tool for a large temperature range of the viscosity of the salt. 
 
 Looking deeper, we notice that the DPMD viscosity values for temperatures in the  range 1400-1600~K slightly underestimate  the experimental data by Tørklep and Øye between $7.8\,\%$ and $15\,\%$.
 A careful observation of the average SACF at these temperatures shows that fluctuations beyond the autocorrelation times of 3~ps is quite significant, therefore introducing a non-negligible error in the integration for the calculation of the viscosity.
 We think that this error can be reduced further by performing more statistically independent simulations at high temperatures. 
 Between 1000~K and 1300~K, the differences with respect to Toerklep and  Øye's  data are between $1.5\,\%$ and $4.5\,\%$. We consider this a very reasonably accurate estimation of the viscosity.
 We have checked that integrating the average SACF function up to a lower maximum autocorrelation time (\(\approx\)~3~ps) produces  systematically lower viscosity values in the range 1000-1400~K and therefore we stay with  the value 7~ps for the maximum integration time of the SACF.

Comparing our DPMD data with the results obtained with simulations of the MACE-MP-0b3 \citep{Batatia2025} foundation model at 1000~K and 1600~K,  we observe that this model overestimates by $11\,\%$ our DPMD results at 1000~K and underestimates the DPMD results at 1600~K by $6.6\,\%$. These results  are presented in magenta circles in Figure~\ref{fig:Viscosity-vs.-temperature} and they are obtained with five thermalized and statistically independent simulations at each temperature. Since the MACE-MP-0b3 foundation model is built from atomic configurations obtained solely with DFT calculations at 0~K, it is not unexpected to find that it does not extrapolate accurately when applied to molten salts at high temperatures.

A comparison of our DPMD data with the results we produce by the use of the MACE-OMAT-MATPES foundation model at 1000~K and 1600~K shows an overestimation of the MACE-OMAT-MATPES model by $7.3\,\%$ at 1000~K and $13\,\%$ at 1600~K. These are quite significant differences. The DPMD viscosity is closer to the experimental data than the MACE-OMAT-MATPES data at 1000~K and further at 1600~K. The reason for this is, once again, the uncertainty of the SACF function at high temperature.

We calculate the viscosity  at 1000~K and 1600~K by using a multihead fine-tuned MACE-OMAT-MATPES foundation model. The differences when compared to the DPMD data is $8\,\%$ and $6.1\,\%$ at 1000~K and 1600~K, respectively. Therefore, we find that the fine-tuning process (with the addition of 64 AIMD configurations specifically for molten MgCl2 salt at 1000~K) improves the out-of-the-box foundation model precision.

As a whole, we find  that the DP model we built from DFT data in the 1000-1600~K temperature range produces viscosity values which are in very  good agreement with experiments and other simulations results, in spite of the sampling  difficulties to obtain accurate SACFs with the Green-Kubo method.
The errors we encounter in the DPMD results as compared to experimental data, can be attributed mainly to the SACF  function sampling more than to a systematic error in the PES of the DP model. 

Finally, we present a comparison with the simulations  performed by Attarian et al. \citep{ATTARIAN2025113409}. Three independent simulations are performed by them at temperatures between 1000~K and 1200~K. As informed by the authors, the viscosity data obtained from their MD simulations, with the use of a model created from PBE-D3 DFT functional data (orange triangles in Figure \ref{fig:Viscosity-vs.-temperature}), underestimate significantly the results obtained by the use of an  R2SCAN DFT functional (violet triangles). The DPMD data we obtain from the model based on the PBE+DFT-D3 functional is in good agreement with the R2SCAN viscosity data by Attarian et al. \citep{ATTARIAN2025113409}, as well as the experimental results.

\subsection{Diffusion coefficient}

\begin{figure}
\begin{centering}
\includegraphics[width=0.98\columnwidth]{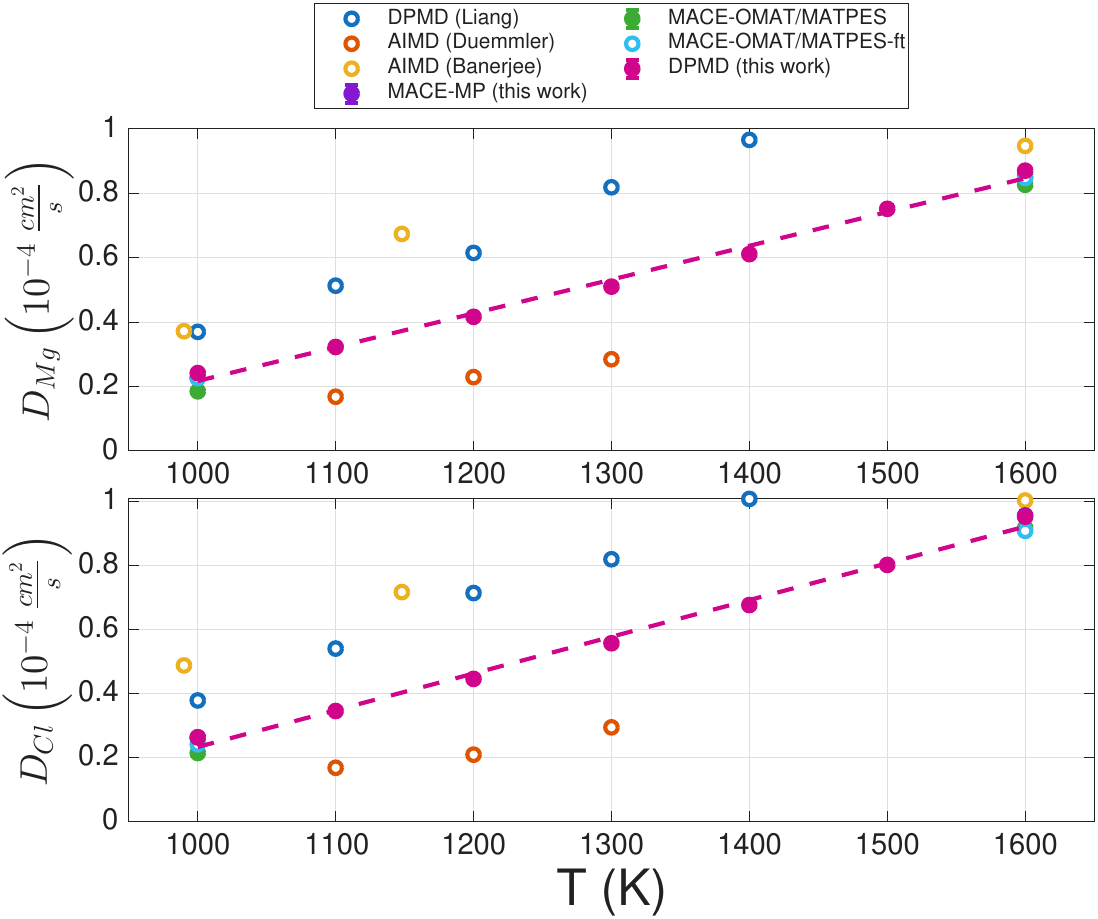}\caption{\label{fig:Diffusion-coefficients-for} Diffusion coefficients for Mg (upper panel) and Cl (lower panel)  in the 1000~K--1600~K range. We compare for each case all the models studied in this work. }
\par\end{centering}
\end{figure}

 We calculate the diffusion coefficients of Mg\textsuperscript{2+} and Cl\textsuperscript{-} ions as a function of temperature. In Figure~\ref{fig:Diffusion-coefficients-for},
we present our results for the DPMD simulations, the MACE-MP-0b3 foundation model and the MACE-OMAT-MATPES foundation model. We also include the multi-head fine-tuned version of the MACE-OMAT-MATPES potential for 1000~K and 1600~K (green open circles). In addition, we compare our results with the simulation results by Duemmler et al. \citep{Duemmler2023} (red open circles), Banerjee et al. \citep{BANERJEE2024125821} (open orange circles), who perform AIMD simulations and Liang et al. \citep{Liang2020b}, who calculate the diffusion coefficients with  DPMD simulations (blue open circles).
The DPMD simulations consistently show that the diffusion of Mg\textsuperscript{2+} ions is smaller than that of Cl\textsuperscript{-} ions in the whole temperature range.
The diffusion coefficient of Cl\textsuperscript{-} ions is larger than Mg\textsuperscript{2+} ions by $6.5\,\%$ to $10.5\,\%$, depending on the temperature. We attribute this difference to the fact that the charge of the Mg\textsuperscript{2+} ions makes them strongly interact  with the first coordination shell of Cl\textsuperscript{-} ions. The MgCl\textsubscript{n} complex  diffuses as a whole until its breakup, creating an arrangement of larger effective mass and therefore lower diffusion coefficient than a single Mg\textsuperscript{2+} ion. The Cl\textsuperscript{-} ions diffuse partially within  this quite stable structure and partially  as individual ions, at a higher diffusion rate, until they are  captured by a MgCl\textsubscript{n} complex.

Using our DPMD results, we calculate the linear fit of the Arrhenius equation, thus obtaining the following results for Mg and Cl, respectively:

\begin{equation}
\log\,\left(D_{Mg}[10^{-4}\frac{cm^{2}}{s}]\right)=\frac{-3393.3}{T[K]}+1.956
\end{equation}
\begin{equation}
\log\,\left(D_{Cl}[10^{-4}\frac{cm^{2}}{s}]\right)=\frac{-3442.9}{T[K]}+2.075
\end{equation}

Our DPMD results are somewhere in the middle between those of Duemmler et al. \citep{Duemmler2023} and those of Liang et al. \citep{Liang2020b}. The fit of the Arrhenius equation by \citep{Liang2020b} for their DPMD results is given by the following equations:

\begin{equation}
\log\,\left(D_{Mg}[10^{-4}\frac{cm^{2}}{s}]\right)_{Liang}=\frac{-3370,28}{T[K]}+2.367
\end{equation}
\begin{equation}
\log\,\left(D_{Cl}[10^{-4}\frac{cm^{2}}{s}]\right)_{Liang}=\frac{-3374.48}{T[K]}+2.425
\end{equation}

The comparison shows that our calculations exhibit a steeper trend
for the diffusion coefficients as a function of temperature. The 
difference in the slopes for the Mg\textsuperscript{2+} ions is about
$0.68\,\%$, while for Cl\textsuperscript{-} ions is about $2.02\,\%$.
However, bigger discrepancies appear in the intercepts, which are
about $21\,\%$ for Mg\textsuperscript{2+} ions and $17\,\%$ for
Cl\textsuperscript{-} ions. These discrepancies in the intercepts imply a smaller diffusion coefficient for both types of ions in our case, when compared to the work of Liang et al., as shown in Figure \ref{fig:Diffusion-coefficients-for}.  This difference can be due to the different Van der Waals energy correction used for the DFT theory. We use DFT-D3 and Liang et al. use DFT-D2 \citep{Liang2020}. A more comprehensive study is needed to clarify this issue.

\begin{figure}

\begin{centering}
\includegraphics[width=0.98\columnwidth]{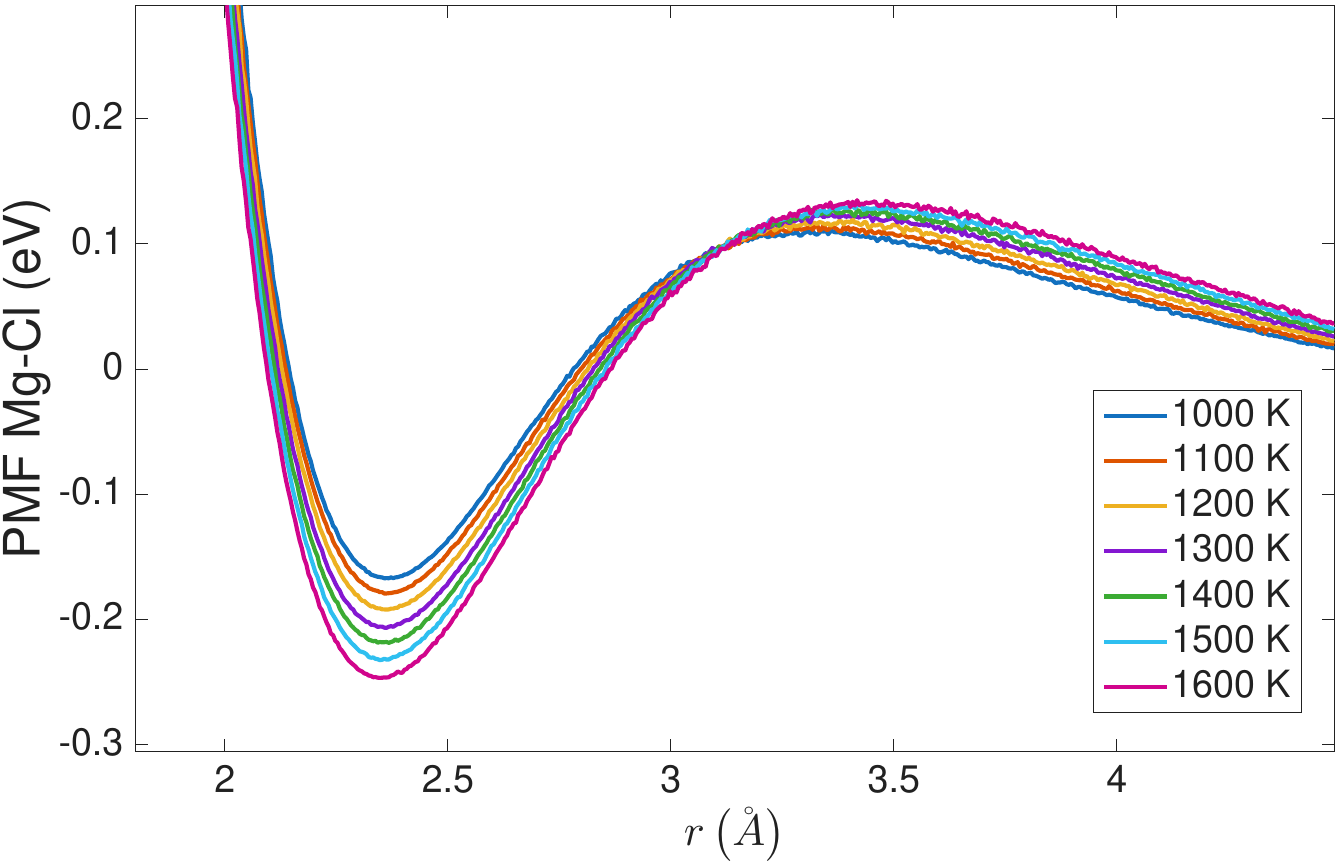}\caption{\label{fig:Potential-Mean-Force}Potential Mean Force for molten MgCl\protect\textsubscript{2}
in the 1000~K--1600~K range.}
\par\end{centering}
\end{figure}

Finally, from the fit of the Arrhenius equation, we calculate the value of the activation
energy for diffusive processes. 
These values are shown in Table \ref{tab:Activation-energies-for}.
This activation values are quite close to the energy needed by a Cl\textsuperscript{-}
ion to leave the first coordination shell around Mg\textsuperscript{2+}
ions. 
The  Potential of  Mean Force for the Mg-Cl interaction is shown  in Figure \ref{fig:Potential-Mean-Force}, as a function of temperature, for the simulations with the DP potential. The energy needed for a Cl\textsuperscript{-} ion to leave the MgCl\textsubscript{n} complex is in the range 0.36--0.38~eV in the studied temperature range. This is  4177~K--4410~K in temperature units. These values are similar to those by Banerjee et al. \citep{BANERJEE2024125821}. We think that the fact that Cl\textsuperscript{-} ions have a higher diffusion coefficient than Mg\textsuperscript{2+} ions indicate  that some of them acquire enough kinetic energy to overcome the effective bonds which form the MgCl\textsubscript{n} complex and, therefore, are able to diffuse independently temporarily in a faster way than the Cl\textsuperscript{-} ions belonging to  the MgCl\textsubscript{n} complex.
According to this physical description of the diffusion process in molten MgCl\textsubscript{2}, the Mg\textsuperscript{2+} ions diffuse with a single rate because they are primarily located at the "center" of the MgCl\textsubscript{n} complexes throughout the salt,
while Cl\textsuperscript{-} ions have a dual diffusion rate, depending on whether they are within a MgCl\textsubscript{n} complex or out of it.
If this dual rate of diffusion were not present, then we would expect the diffusion coefficients of both types of ions to be very similar, because they would both diffuse as a part of a the MgCl\textsubscript{n} complex.


\begin{table}

\begin{centering}
\begin{tabular}{|c|c|c|}
\hline 
 & E\textsubscript{a }(eV) & Error (eV)\tabularnewline
\hline 
\hline 
Mg\textsuperscript{2+} & 0.292 & 0.005\tabularnewline
\hline 
Cl\textsuperscript{-} & 0.297 & 0.005\tabularnewline
\hline
\end{tabular}
\caption{\label{tab:Activation-energies-for}Activation energies for molten MgCl\protect\textsubscript{2} ions. }
\par\end{centering}
\end{table}


\section{Accuracy and performance of the MLIPs}
\label{sec:performance_MLIPs}

We report on the performance of the different MLIPs employed throughout this work, which is an important aspect to make an  appropriate choice for a given system.
First, we  discuss a performance metric for the simulations. The AIMD and the DPMD simulations are done in a single Intel\textsuperscript{\textregistered} Xeon\textsuperscript{\textregistered} CPU E5-2650 v4 node with 24 cores.
The MACE simulations are run in a single node with Intel\textsuperscript{\textregistered} XeonIntel\textsuperscript{\textregistered} Max 9462 CPU with 64 cores and a single Intel Max series GPU (with a memory of 128GB)

Table \ref{tab:accuracy-performance} shows  the performance of AIMD, DPMD and MACE simulations. Reference speed-up values for our simulations are also provided. The speed-up is normalized to one for AIMD simulations, as it is the  method of highest accuracy and highest computational cost.

We observe  that the DPMD simulations greatly reduce the computational cost with respect to AIMD simulations. A \(\approx4000\times\) speed-up was typically found for the  DPMD simulations for molten MgCl\textsubscript{2} salt. This significant speed-up allows longer MD trajectories  to perform 
the calculation of ion diffusion, viscosity and thermal conductivity  with good accuracy within reasonable computing times. These properties would be very challenging to obtain with AIMD.
Given that the accuracy of these models, benchmarked against DFT data is very good, we think that DP models are a suitable tool to study  pure and mixed  molten salts.

The speed-up of MACE simulations relative to AIMD simulations is about $60\,\%$ of the DPMD speed-up. It is a significant improvement over DFT calculations. Even more so, if we consider that these foundation models span most of the Periodic Table \citep{Batatia2025}. These models are very fast compared to AIMD and produce results which agree with AIMD simulations in many thermophysical properties. However,  there are a few discrepancies that are worth mentioning such as Mg-Mg RDF and viscosity (see Figures \ref{fig:MgCl2-RDFs-1000K} and \ref{fig:Viscosity-vs.-temperature}). 
Benchmarking  against DFT is very important and fine-tuning is usually 
mandatory to achieve good results.
Fine-tuned MACE models  follow essentially the same performance metrics as the original foundation models, while improving on the physical properties  that  the out-of-the-box model give poorly.
It should also be mentioned that MACE models are not suitable for use on CPU-only HPC systems, as the non-linear terms of the neural network make a strong use of GPU processing. We have observed very slow performances on these last cases, comparable in performance metrics to AIMD simulations. 

\begin{table}
\begin{centering}
\begin{tabular}{|c|c|c|c|}
\hline 
\parbox{2cm}{\centering Simulation type} & 
\parbox{2cm}{\centering Number of atoms} & 
\parbox{2.5cm}{\centering Performance (katom-step/s)} & 
Speed-up \tabularnewline
\hline 
\hline 
AIMD & 108 & 0.0107 & 1\tabularnewline
\hline 
DPMD & 4320 & 43.1396 & 4023\tabularnewline
\hline 
MACE & 4320 & 26.2202 & 2445\tabularnewline
\hline 
\end{tabular}
\caption{\label{tab:accuracy-performance}Performance metric for AIMD, DPMD and MACE simulations.}
\par\end{centering}
\end{table}

\section{Discussion and Conclusions}

We have investigated the structural, thermodynamic and transport properties of molten MgCl\textsubscript{2} over a wide temperature range by combining Ab Initio molecular dynamics, Deep Potential Molecular Dynamics, and several foundation machine-learning interatomic potentials, including both pre-trained and fine-tuned models.

The AIMD simulations provide an accurate description of the structural and thermophysical properties of molten MgCl\textsubscript{2} despite the relatively small simulation cell (108 atoms) and the limited trajectory length. Whenever statistical convergence is achieved, the calculated properties are in good agreement with available experimental measurements and previous computational studies. In particular, the density, radial distribution functions, coordination numbers and angular distribution functions are consistently reproduced, indicating that the chosen AIMD setup provides a reliable description of the liquid structure.

The structural analysis further shows that the local ionic arrangement remains remarkably stable throughout the investigated temperature range. The radial distribution functions exhibit only minor temperature-dependent changes, indicating that strong ionic interactions dominate over thermal disorder. Likewise, the angular distributions remain highly distorted but essentially temperature independent, suggesting that the local coordination environment is largely preserved upon heating.

The larger simulation times made possible by ML-based molecular dynamics allow the calculation of transport properties that are prohibitively expensive to obtain directly from AIMD. The calculated heat capacity at constant volume exhibits only a mild decrease with increasing temperature, while the thermal conductivity shows the expected moderate reduction throughout the studied range. These results contribute reliable reference data for pure molten MgCl\textsubscript{2}, which are valuable for future investigations of molten-salt mixtures and technological applications.

The calculated viscosities are in good agreement with the experimental values of Tørklep and Janz, although some discrepancies remain. Since viscosity is obtained through the Green--Kubo formalism, a more extensive sampling of the stress autocorrelation function would likely improve the statistical convergence and further increase the accuracy of the calculated values.

The diffusion coefficients we obtain from DPMD display the expected increase with temperature and lie within the range reported in previous experimental and computational studies. As expected, Mg\textsuperscript{2+} ions diffuse more slowly than Cl\textsuperscript{-} ions over the entire temperature interval because of their stronger electrostatic interactions with the surrounding liquid. The comparison with the MACE foundation models shows very good agreement  at the lowest and highest temperatures considered, suggesting that similar agreement is expected throughout the intermediate temperature range.

From a methodological perspective, this work demonstrates that machine-learning interatomic potentials provide an effective route for extending the length and time scales accessible to first-principles simulations. A neural-network DP potential trained from scratch using representative AIMD configurations accurately reproduces the reference energies and forces, while enabling simulations that are orders of magnitude larger and longer than those feasible with AIMD directly.

The trained DP potential also exhibits a high degree of robustness. Models initialized from different random neural-network parameters converge to essentially the same energy and force predictions, indicating that the resulting potential energy surface is largely independent of the specific optimization path. This robustness enhances the reproducibility of the model and eliminates the need for repeated retraining using different network initializations.

Our results further suggest that relatively compact invariant neural-network architectures are sufficient to reproduce bulk properties of molten MgCl\textsubscript{2}. Increasing the network depth or width is therefore unlikely to provide substantial improvements for homogeneous liquid systems. Whether similar architectures remain adequate for chemically more complex environments, such as multicomponent molten salts or solid--liquid interfaces, remains an open question.

Future work will extend the present methodology to molten-salt mixtures with varying compositions and to interfacial systems involving molten salts in contact with crystalline materials, particularly iron surfaces. Such studies are of considerable technological interest because they are directly related to corrosion phenomena in advanced molten-salt energy systems. We must make clear that molten-salt mixtures, systems with metal/molten salt interfaces and corrosion studies will require the extension of models like the one developed from scratch in this work. New training data will be necessary in order to take into account interactions which are not present in the AIMD dataset used for model training in this publication. Consequently, a validation of this new model by comparison to DFT data will also be necessary. Furthermore, it is interesting to see how out-of-the-box foundation models behave in these scenarios.

\section*{Conflict of interests}

The authors have no conflicts of interest to disclose.

\section*{Supplementary Material}

The Supplementary Material file  contains complementary results  and deeper explanations on the following topics: RDFs and coordination numbers of Mg-Cl, Mg-Mg and Cl-Cl ion pairs (Section I), details of the thermal conductivity  (Section II) and viscosity (Section III) calculations. It includes also a section with accuracy checks of the deep potential developed in this study (Section IV).

\section*{AI use statement}
 
 Artificial intelligence (AI) tools such as ChatGPT (openAI) and Claude, version Sonnet (Anthropic) are used  to improve readability, grammar, and clarity of the manuscript. The complete text is fully revised and corrected  by the authors. AI is also used to improve performance on some postprocessing scripts, with full verification by the authors of  the  final version.
 
 The scientific decisions and analyses, interpretation of the results, and all conclusions are performed by the authors, who take full responsibility for the content of this manuscript.

\section*{ Acknowledgments }

 C. P., M.A.B and V.L.V. thank CONICET for partial support of this work through grants PIP1220210100546CO and PIP11220220100330CO. CNEA  is also gratefully acknowledged for providing  supercomputing resources for our LabSim Supercomputing Lab. The authors also thank SICyT and the sys-admins for support and computing time in Clementina XXI (projects PCI-112 and PAD-148, 2025). R. L. and C. P. thank Pablo Piaggi for fruitful discussions on the accuracy check  of the deep potential with a committee of models. 

 \section*{CRediT authorship contribution statement}

 {\bf Roberto Llovera:} Conceptualization, Data curation, Investigation, Methodology, Software, Validation, Formal Analysis
 Visualization, Data Curation Writing original draft, reviewing  and editing.
 
 {\bf María Andrea Barral and Verónica Vildosola:} Conceptualization, Investigation, DFT calculations, Validation, Methodology, Formal Analysis, Resources Writing (review and editing), Funding Acquisition.
 
 {\bf Florencia Cantargi:} Conceptualization, Funding Acquisition, Resources, Supervision, Project administration. 
 
 {\bf Claudio Pastorino:} Conceptualization, Data curation, Investigation, Methodology, Software, Validation, Formal Analysis, Writing original draft, reviewing  and editing, Resources, Funding Acquisition, Supervision.

 \section*{Data availability}

 The data that support the findings of this study are available from the corresponding  authors upon request.
 
\bibliographystyle{apsrev4-1}
\bibliography{bibtex}
\end{document}